\documentclass[%
 reprint,
 amsmath,amssymb,
 aps,
]{revtex4-2}
\usepackage{makecell}
\usepackage{nicefrac}
\usepackage{booktabs}
\usepackage{graphicx}
\usepackage{dcolumn}
\usepackage{bm}

\usepackage{color}
\usepackage[dvipsnames]{xcolor}
\definecolor{myblue}{RGB}{0, 0, 0}
\definecolor{darkBlue}{RGB}{0, 0, 0}
\newcommand{\review}[1]{ {\color{myblue} #1 } }

\begin{document}


\title{Topological Field Theory and Stochastic Dynamics}

\author{Igor V. Ovchinnikov}
\email{igor.vlad.ovchinnikov@gmail.com}
\affiliation{ Device Research Laboratory, Department of Electrical Engineering, University of California at Los Angeles, Los Angeles, 90095, CA, USA}
\altaffiliation[present address:]{R\&D, CSD, ThermoFisher Scientific Inc., 200 Oyster Point, South San Francisco, 94080, CA, USA}

\date{\today}

\begin{abstract}
In the late 1980s, Baulieu and Grossman demonstrated that the supersymmetric formulation of Langevin stochastic differential equations (SDEs), proposed earlier by Parisi and Sourlas, belongs to the family of Witten-type topological field theories (TFTs). From a certain angle, this finding may appear puzzling: TFTs have no local degrees of freedom, whereas SDEs do exhibit local fluctuations. In this paper, we address this apparent contradiction in the context of the supersymmetric theory of stochastic dynamics, a generalization of the Parisi–Sourlas-Baulieu-Grossman approach to SDEs of arbitrary form. The resolution lies in recognizing that, as a mathematical construct, each SDE encodes two subsystems: the continuous-time dynamical system (DS) and its noise. The Parisi–Sourlas construction employs a gauge-fixing procedure to rewrite the partition function of the noise (PFN) in terms of the variables of the DS, resulting in the Witten index. Being a topological invariant, this representative of the PFN is independent of the duration of the evolution, reflecting the absence of dynamics in the noise, as expected since the noise is a static probabilistic entity that experiences no backaction from the DS. The partition function of the DS itself is obtained by twisting, i.e., by imposing anti-periodic boundary conditions (APBC) on the Faddeev–Popov ghosts, thereby rendering them physical -- a step that can be justified by interpreting the ghosts, e.g., as differentials used in numerical experiments to track Lyapunov exponents. Since APBC are not compatible with the topological character of the Witten index, one might be tempted to conclude that the twist destroys the topological/BRST supersymmetry (TS) of the model altogether. In the context of stochastic dynamics, however, this conclusion is incorrect. The TS remains intact because it is a property of the stochastic evolution operator (SEO), the fundamental construct corresponding to open boundary conditions and known in DS theory as the generalized transfer operator. The Witten index and the partition function do not define different models; rather, they are different objects of the same model defined by the SEO. What APBC do change, however, is that the partition function exhibits exponential growth in time when TS is spontaneously broken -- a situation that, as it turns out, is equivalent to the definition of chaos in random DSs introduced by Ruelle. \review{Within the resulting picture of chaos, 1/f noise emerges as a consequence of the Goldstone theorem, while the butterfly effect is described by an effective field theory that may possess a hidden topological structure, as we speculate on the basis of the Ginzburg-Landau approach and AdS/CFT duality.}
%
\end{abstract}

\maketitle

\section{Introduction}
A decade after the introduction of the supersymmetric formulation of Langevin stochastic differential equations (SDEs) by Parisi and Sourlas~\cite{Parisi_Sourlas_1979,Parisi_Sourlas_1982}, Baulieu and Grossman~\cite{Baulieu_1988,Baulieu_1989} showed that this construction belongs to the class of Witten-type, or cohomological, topological field theories (TFTs)~\cite{Witten_1982,CecottiGirardello1982,CecottiGirardello1983,CecottiGirardello1984, TFT_BOOK,Labastida_1989,Baulieu_1988,Baulieu_1989,Baulieu_1989_1,Witten_1988,Witten_1988_1,Frenkel_2007,Niemi_1986,Losev_2005,LabastidaMarino_2005_book_TQFT,Schwarz_1993_book_QFTTopology,Sadovski_2025_TFTs,labastida_1997_lectures}. 
More recently, this result has been generalized from Langevin dynamics~\cite{Cardy_1985,ZinnJustin_1986,Olenskoi_2004,Dijkgraaf_2010,Drummond_2012,Kaviraj_2022,Rychkov_2023,Nakayama_2025,Nicolis_2026,Le_Floch_2025,Kaviraj_2020,Kaviraj_2021} and a few other specific SDEs \cite{Gozzi_1990,Gozzi_1994_Lyapunov,Deotto_2003,Tailleur_2006,Kleinert_Shabanov_1997} to SDEs of arbitrary form~\cite{Ovchinnikov_2016,Ovchinnikov_2016_1,Ensslin_2016,Sethi_Weiderpass_2026} -- the most general class of continuous-time dynamical systems with noise whose applicability extends beyond physics and related disciplines. Within this broader framework, which may be referred to as the supersymmetric theory of stochastic dynamics (STS),~(stochastic) chaos~\cite{Ruelle_2014,Gilmore_1988,Gilmore_Lefranc_2002_book,Eckmann_Ruelle_1985_RMP,Devaney_1992_book,Strogatz_2017_book,Mangiarotti_2021,Yorke_1975,Tel_2015,Elaskar_2023,Baxendale_1986,Arnold_1986,Kapitaniak_1990,Uthamacumaran_2021} is identified as the spontaneous breakdown of topological or Becchi–Rouet–Stora–Tyutin (BRST) supersymmetry (TS), while the Goldstone theorem provides a remarkably simple yet robust explanation of $1/f$ noise~\cite{Voss_1979,Dutta_1981,Keshner_1982,Wang_2010,Szendro_2001,Pettersen_2014,Biology1fNoise,BookHeartBrainNoise,Preis_2011,Bedard_2006} -- an experimental signature of chaos that has long resisted a satisfactory theoretical account within more conventional approaches. More broadly, this framework reverses the traditional view of chaos as a form of dynamical randomness, suggesting instead that it conceals an underlying long-range order, potentially of topological character~\cite{Ovchinnikov_2024,Ovchinnikov_2026}.

TFTs are often characterized as models without local degrees of freedom. For this reason, it is not surprising that TFTs provide effective low-energy descriptions of certain condensed-matter systems in which local excitations are gapped and, at sufficiently low temperatures, fluctuations can be neglected (see, e.g., Refs.~\cite{Altland_Simons_book_2010,Qi_Zhang_2011,Hansson_2004} and Refs therein)~\footnote{It must be pointed out that TFTs describing low-temperature condensed matter phases belong to Schwarz-type or quantum TFTs -- the other major class of TFTs~\cite{TFT_BOOK}.}. For the same reason, the idea that TFTs might describe SDEs -- where local fluctuations are intrinsic -- may appear puzzling. 

More specifically, due to the presence of TS, only zero-eigenvalue supersymmetric singlets contribute to the functional proposed by Parisi and Sourlas as the partition function of the model; the contributions of all nonsupersymmetric states cancel out. As a result, this functional is a topological invariant -- the Witten index -- independent, in particular, of the duration of evolution, as if the model had no excitations. This may invite the interpretation that all nonsupersymmetric states are unphysical, which is in line with logic of BRST quantization in quantum field theory (QFT)~\cite{Peskin_Schroeder_1995_book}.

In the context of stochastic dynamics, however, declaring all eigenstates with non-zero eigenvalue unphysical does not appear to be a viable option. Indeed, these states are essential, in particular, for describing out-of-equilibrium Fokker–Planck evolution. 

In this paper, we discuss the resolution of this tension. It begins by recognizing that the definition of an SDE involves two distinct subsystems: the noise and the continuous-time dynamical system (DS) affected by it. By construction, the Parisi–Sourlas functional is essentially the partition function of the noise, albeit formulated in the variables of the associated DS. From this perspective, its independence from the duration of evolution is natural: it reflects the absence of excitations in the noise, which, by definition, is a static probabilistic entity that does not experience backaction from the DS.

The partition function of the DS itself is recovered via the twist, i.e., by imposing antiperiodic boundary conditions on the Faddeev–Popov ghosts. In the language of QFT, this amounts to declaring the ghosts physical, a step that requires a consistent interpretation for the ghosts. Such an interpretation does exist: the ghosts can be viewed as infinitesimal differentials used in numerical experiments to track Lyapunov exponents~\cite{Graham_1988,Gozzi_1994_Lyapunov,Ovchinnikov_2024}.

The twist is not compatible with the topological character of the Parisi–Sourlas functional. This fact may suggest that the twist destroys the TS of the model altogether. In the context of stochastic dynamics, however, this conclusion would also be misleading. The key point is that the fundamental object governing the dynamics is the stochastic evolution operator (SEO) -- the noise-averaged pullback induced by the SDE-defined diffeomorphisms, known in the random DSs theory as the generalized transfer operator~\cite{Ruelle_2002}. The twist does not affect the SEO which corresponds to open boundary conditions. Consequently, it does not alter the presence of TS when the latter is understood as a property of the SEO. Incidentally, the spontaneous breakdown of the TS of the SEO is equivalent to the definition of chaos in random DS theory proposed by Ruelle~\cite{Ruelle_2002,Ovchinnikov_2016}.

With the SEO as a central concept for SDEs, it may be useful to review how it relates to other objects that arise in stochastic dynamics. Alongside the above discussion of the interpretation of the STS, providing such a synthesis is another goal of this paper, summarized in Table.\ref{tab:Objects}. \review{We also qualitatively discuss the problem of constructing effective field theories (EFT) of chaotic dynamics. Using a Ginzburg–Landau-like approach together with insights from the AdS/CFT correspondence, we speculate that some such EFTs may possess a hidden topological structure.}

The paper is organized as follows. In Sec.~\ref{sec:DS_perspective}, we introduce the model from the perspective of random dynamical systems theory. In Sec.~\ref{Sec:Weighted_Traces}, we review the Parisi–Sourlas construction and its twisted version, along with their relation to the traditional theory of SDEs. In Sec.~\ref{Sec:Physicality}, we discuss the physicality of non-supersymmetric states and the corresponding interpretation of matrix elements of the SEO. In Sec.~\ref{Sec:StateToState}, we examine instantons and the generating functional. \review{In Sec.\ref{Sec:EFT}, we discuss 1/f noise and Ginzburg-Landau approach to EFT and speculate about the possibility that some chaotic models may contain non-trivial topological spatio-temporal structure. Section~\ref{Sec:Conclusion} concludes the paper.}

\section{Stochastic evolution of differential forms}
\label{sec:DS_perspective}
In this section, we introduce stochastic dynamics from the perspective of random DS theory \cite{Ruelle_2002}. Compared to the path-integral representation traditionally used in the literature on the Parisi-Sourlas construction, this viewpoint offers at least two advantages: it highlights the fundamental mathematical meaning of the SEO, and it is free from the It\^o-Stratonovich ambiguity inherent in the path-integral representation of evolution operators.
\subsection{The model}

We focus on the following class of 1+0 models, which may as well represent lattice versions of models with higher-dimensional base-spaces,
\begin{subequations}
\label{Eq:SDE}
\begin{eqnarray}
    \dot \varphi(t) = f(\varphi(t)) + (2\Theta)^{1/2}g_a(\varphi(t))\xi^a(t) \equiv F(\varphi,t),\label{Eq:SDE_only}
\end{eqnarray}
where $t\in(0,T)$ is time with $T$ being duration of evolution, $\varphi\in X$ is the dynamical variable of the DS from the phase/state space which is assumed a $D$-dimensional, compact, orientable, smooth manifold, $X$, $f\in TX$ is a vector field representing the deterministic law of evolution, $\Theta$ is noise intensity, $\xi^a\in\mathbb{R}$ and $g_a\in TX, a=1,...,N, N\ge D$ is a set of noise variables and a set of vector fields that specify how the noise is coupled to the system. For simplicity, the noise is assumed Gaussian white. Its (normalized) partition function is given as,
\begin{eqnarray}
    Z_{N} = \iint D\xi P_N(\xi)  = 1,\; P_N(\xi) \propto e^{-\int_0^T \xi(\tau)^2 d \tau/2},\label{Eq:PartitionFunctionNoise}
\end{eqnarray}
\end{subequations}
where $P_N$ is the normalized probability distribution. The fundamental average of the noise is given as,
\begin{eqnarray}
   \overline{\xi^a(t)\xi^b(t')}=\delta^{ab}\delta(t-t'),
\end{eqnarray}
where we introduced the notation
\begin{eqnarray}
    \overline{ A } = \iint A(\xi) P(\xi)D\xi,\label{Eq:definitionOfNoiseAveraging}
\end{eqnarray}
with $A(\xi)$ being some functional of $\xi$.

There are no restrictions on $f$ and $g$ beyond sufficient smoothness (see below). Accordingly, Eq.~(\ref{Eq:SDE}) generalizes Langevin SDEs considered by Parisi and Sourlas, corresponding to $f = -\nabla U$, $g = \mathrm{const}$, and $X = \mathbb{R}^D$. Anticipating the discussion in the following section, the key qualitative difference introduced by this generalization is the possibility of spontaneous breakdown of topological supersymmetry, that is, Eq.~(\ref{Eq:SDE}) can exhibit chaotic behavior, whereas Langevin SDEs are never chaotic.

\begin{figure}[t] 
    \centering
\includegraphics[width=0.99\linewidth,height=4.5cm]{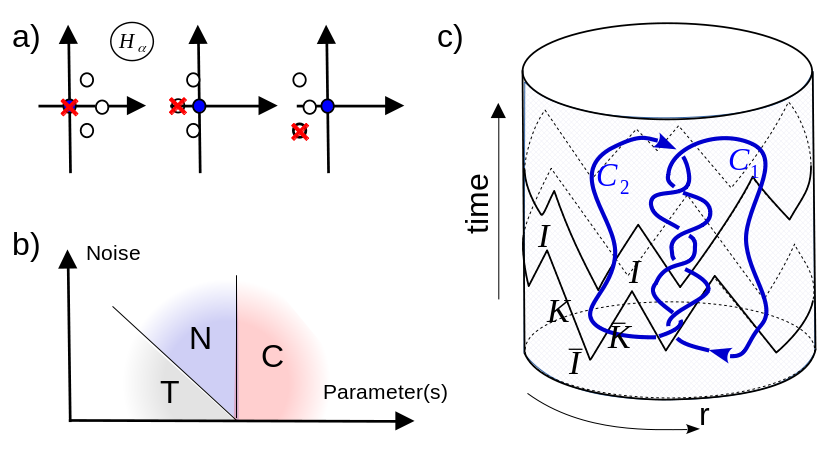}
\caption{\review{\label{Fig_1} {\bf (a)} The three types of SEO spectra corresponding to: \emph{(left)} unbroken TS (or BRST symmetry), when the ground state (red cross) is the zero-eigenvalue supersymmetric state (blue circle at the origin); \emph{(center)} broken TS, when the ground state has a negative eigenvalue, which is known in dynamical systems theory as the “pressure” of chaotic dynamics; and \emph{(right)} the case when both TS and the pseudo-time-reversal symmetry of the SEO are broken by a nonsupersymmetric ground state with complex eigenvalue. {\bf (b)} At low noise intensity, there are two types of chaos: conventional chaos, or the C-phase, where TS is broken at the Gaussian level by the fractal structure of the (strange) attractor; and noise-induced, or instantonic, chaos, or the N-phase, where TS is broken by noise-induced instantons. In the deterministic limit, the N-phase collapses onto the boundary between the C-phase and the symmetric phase marked by T. {\bf (c)} In the N-phase of the overdamped sine-Gordon equation with nonpotential driving \cite{Ovchinnikov_2024}, the dynamics is dominated by (anti-)instantonic processes ($I$ and $\bar I$) of (creation) annihilation of pairs of left- and right-moving kinks and antikinks ($K$ and $\bar K$). The solitons move at a fixed velocity, suggesting that the EFT is a Lorentzian CFT. As we speculate in Sec.~\ref{Sec:EFT_TFT}, the holographic $AdS_3/CFT_2$ dual of the EFT may allow computation of $GL(1|1)$ Wilson loops related to the Alexander polynomial~\cite{Rozansky_1992}.}}
\end{figure}

\subsection{Stochastics vs. Statistics}
\label{Sec:Stochastics_Statistics}

Let us now pause for a moment to examine more closely the structure of the dynamics defined by Eqs.(\ref{Eq:SDE}). The first point to note is that the model consists of two distinct components: the time-dependent noise and the continuous-time, non-autonomous DS affected by it. The noise is purely probabilistic and possesses no intrinsic dynamics. Moreover, there is no backaction from the DS onto the noise. This reflects the fact that the noise represents an environment -- a much larger DS coupled to the DS under consideration.


Unlike the noise, which is defined statistically through its partition function, the DS is specified via its equations of motion (EoM). 
These equations can be viewed as a relation between the DS and the noise variables. It is therefore tempting to consider a change of variables in the functional integral of Eq.~(\ref{Eq:PartitionFunctionNoise}), replacing the integration over \(\xi\) with an integration over \(\varphi\), with the aim of constructing the partition function of the DS.

There are two reasons to exercise caution in using this approach. First, even if the change of variables were carried out successfully, it would not yield the partition function of the DS itself. Rather, it would reproduce the partition function of the noise expressed in terms of DS variables. As discussed above, the DS and the noise are distinct dynamical systems, and their partition functions must therefore also be different. As it will be established later, this is indeed the case.

Second, the dynamics of the DS is richer than the statistics of the noise. Indeed, each trajectory corresponds to a unique noise configuration, but the converse is not true: for a given noise configuration, there are infinitely many possible trajectories because the boundary conditions in EoM for $\varphi$ are open, that is, unspecified. In other words, the DS contains more degrees of freedom than the noise~\footnote{This is particularly clear in the lattice version of the model. There, the DS variables are defined at $N+1$ discrete time points, $t_k = k \Delta t$, $k=0,\dots,N$, with $\Delta t = T/N$ and $N \gg 1$, giving $(N+1)D$ DS variables in total. The noise, in its turn, acts during the transitions between the neighboring time points, effectively living on a dual lattice, and thus has only $N D$ variables.}.

This mismatch can be resolved by imposing boundary conditions on $\varphi$. In what follows, we discuss that different choices of boundary conditions lead to different objects, each with its own meaning.


\subsection{Stochastic Evolution Operator}
\label{Sec:SEO}

In fact, Eq.~(\ref{Eq:SDE}) does not naturally belong to the domain of statistics or statistical physics, and the partition function is not the most fundamental object governing the model. Rather, SDEs \cite{Kunita_1997_book_SDEs,Le_Jan_Watanabe_1984,Heirer_2018} naturally reside within the framework of random DS theory~\cite{Arnold_1988,Pesin_1977,Baxendale_1986,Arnold_1986,Arnold_1983,Bedrossian_2022_LyapunovExpReview,Ruelle_2002}. From the DS theory perspective, the central object is the transfer operator -- the analogue of the evolution operator in quantum theory. Partition-function-like quantities, or weighted traces as they are called sometimes, can be obtained from this operator.

Of particular interest is the generalized transfer operator~\cite{Ruelle_2002}, which, in the context of continuous-time dynamics, can be referred to as the stochastic evolution operator (SEO). In terms of algebraic topology, the SEO is a representation of dynamics generated by SDE on the exterior algebra, $\Omega$, i.e., the space of all differential forms,
\review{\begin{eqnarray}
    &\psi(\varphi,\chi) = \sum\nolimits_{k=0}^D \psi^{(k)}_{i_1...i_k}(\varphi)\chi^{i_1}...\chi^{i_k} \in \Omega^\bullet_\varphi(X).
\end{eqnarray}
}Here, $\psi^{(k)}$'s is an antisymmetric tensor and $\chi^j\sim d\varphi^j\wedge, $ are antisymmetric differentials, $\chi^i\chi^j=-\chi^j\chi^i$, which, from the point of view of QFT, are fermionic fields that will later be identified as Faddeev-Popov ghosts. Borrowing terminology from quantum theory, we refer to $\psi$'s and $\Omega$ as wavefunctions and the Hilbert space, respectively.

To define SEO, let us first "solve" the SDE and introduce the family of maps, $M_{tt'}(\xi):X\to X$, such that a trajectory corresponding to initial condition $\varphi_0$ at $t'$ and noise-configuration $\xi$ is given as $\varphi(t)=M_{tt'}(\xi)(\varphi_0)$. Provided that $f$ and $g$'s in Eq.(\ref{Eq:SDE}) are sufficiently smooth in $X$, the maps are diffeomorphisms even for noise configurations that are not smooth in time~\cite{Slavik_2013}. This means that two infinitesimally close initial points remain close for finite-time evolution~\footnote{In the limit of infinitely long evolution, close trajectories can part signaling the presence of the butterfly effect.}.

The SEO is defined as (see Ref.\cite{Ovchinnikov_2016} for details),
\begin{eqnarray}
&\hat M_{T0} = \overline{M_{0T}^*} = e^{-T\hat H},  \label{Eq:SEO_operators}
\end{eqnarray}
where $M_{0T}^*$ is the pullback induced by the corresponding noise-configuration-dependent diffeomorphism on the wavefunction, the bar denotes averaging over the noise in Eq.(\ref{Eq:definitionOfNoiseAveraging}), and the \emph{infinitesimal} SEO \footnote{In the traditional theory of SDEs, this operator is known as the evolution operator in the Stratonovich interpretation of SDEs.\cite{Kunita_1997_book_SDEs,Le_Jan_Watanabe_1984}}  \footnote{Here we introduce summation over the repeated indices.}
\begin{eqnarray}
\label{Eq:SEO_LieDerivatives}
 &\hat H = \hat L_{f} - \Theta \hat L_{g_a}\hat L_{g_a}.
\end{eqnarray}
\review{Here, $\hat L$'s are Lie derivatives, e.g.,
\begin{eqnarray}
   \hat L_f = f^i \partial / \partial \varphi^i   +  f^i_{'j}\chi^j \partial/\partial \chi^i  ,
\end{eqnarray}
that can also be given via the Cartan formula,}
\begin{eqnarray}
   \hat L_f = [\hat d, \hat \imath_f],
    \hat d = \chi^i \partial/\partial \varphi^i, \; \hat \imath_f = f^i \partial/\partial \chi^i,
\end{eqnarray}
with $\hat d$ being the exterior derivative, $\hat \imath$ being the interior multiplication, and square brackets denoting bi-graded commutator, $[\hat A, \hat B] = \hat A\hat B - (-1)^{\deg \hat A \deg \hat B}\hat B\hat A $. 

The SEO is $\hat d$-exact,
\begin{eqnarray}
    \hat H = [\hat d, \hat {\bar d}],\; \hat {\bar d} = \hat \imath_f - \Theta \hat \imath_{g_a}\hat L_{g_a},\label{Eq:DExactH}
\end{eqnarray}
as follows from nilpotentcy of exterior derivative, $\hat d^2=0$, and the fact that a commutation with $\hat d$ is a differentiation, $[\hat d, \hat A \hat B ] = [\hat d, \hat A ] \hat B + (-1)^{\deg \hat A} \hat A [\hat d, \hat B ]$. 

Evolution operators of this form are specific to Witten-type TFTs (see, e.g., Ref.\cite{TFT_BOOK} and refs therein) such as topological quantum mechanics~\cite{Losev_2005,Faddeev_1993,Freed_1999}, and $\hat d$ can be recognized a topological supersymmetry (TS), which is not surprising -- the exterior derivative is the most fundamental concept in algebraic topology~\cite{Nakahara_book}. In Sec.~\ref{Sec:TFT}, we will discuss that STS belongs to this family of models.

In the next section, the path-integral representation of the SEO will also be useful. It can be introduced as
\begin{eqnarray}
    \hat M_{T0} = \iint_\text{open BC} e^{S(\Phi)}D\Phi .\label{Eq:SEO_operator}
\end{eqnarray}
Here, the functional integration is carried out over paths with open boundary conditions (BC), that is, over paths that connect the in- and out- arguments of the SEO, $\Phi=\varphi,B,\chi,\bar\chi$ is a collection of fields containing the original fields $\varphi,\chi$ and their momenta $B,\bar\chi$, and the action can be defined as
\begin{subequations}
\label{Eq:Action}
\begin{eqnarray}
    S(\Phi) = \{\mathcal{Q},\bar {\mathcal{Q}}\}.\label{Eq:STS_Action}
\end{eqnarray}
Here, the notation in the r.h.s. denotes bi-graded commutator, $\{A,B\} = AB - (-1)^{\deg A \deg B} BA$, 
\begin{eqnarray}
    \mathcal{Q} = \int_{0}^T d\tau \left(\chi^i {\delta}/{\delta \varphi^i} + B_i{\delta}/{\delta {\bar\chi}_i}\right),\label{Eq:SUSY}
\end{eqnarray}
and
\begin{eqnarray}
    \bar {\mathcal{Q}} = \int_{0}^T d\tau \left(i\bar\chi_i \dot \varphi^i- \bar d\right),\label{Eq:QBar}
\end{eqnarray}
\end{subequations}
is what can be identified as a gauge fermion with $\bar d = i\bar\chi_i (f^i - \Theta g^i_a L_{g_a})$ and $L_{g_a} = \{\mathcal{Q},i\bar\chi_ig_a^i\}$ being the path-integral versions of the operator $\hat {\bar d}$ from Eq.(\ref{Eq:DExactH}) and Lie derivative, respectively.

\color{myblue}
Likewise, $\mathcal Q$ is the path-integral version of TS. In the gauge-fixing interpretation of STS (see Sec.~\ref{Sec:ParisiSourlasConstruction}), however, this object can be identified with BRST symmetry. In the literature on cohomological TFTs, both terms -- TS and BRST symmetry -- are used. At the same time, as discussed in Sec.~\ref{Sec:ParisiSourlasConstruction}, there are reasons why TS may be a more preferable designation for STS. Accordingly, we will primarily use the term TS.

\subsubsection{It\^o-Stratnovich dilemma}

The r.h.s. of Eq.~(\ref{Eq:SEO_operator}) has an intrinsic ambiguity, common to all path-integral representations of evolution operators. In the traditional theory of SDEs, this ambiguity is known as the It\^o–Stratonovich dilemma. The equality in Eq.~(\ref{Eq:SEO_operator}) effectively resolves this ambiguity~\cite{Ovchinnikov_2025}, since the SEO on the l.h.s -- defined in Eq.~(\ref{Eq:SEO_operators}) -- is unique and corresponds to the Stratonovich interpretation of SDEs. 

Equivalently, the It\^o–Stratonovich dilemma amounts to the ambiguity in operator ordering when passing from the path-integral to the operator representation of the evolution operator (see, e.g., Sec.~3.3 of Ref.~\cite{Ovchinnikov_2025}). In particular, a product $B\varphi$ appearing in the path integral is represented at the operator level as $(1-\alpha)\hat B \hat \varphi + \alpha \hat \varphi \hat B$, with $\alpha=0$ and $\alpha=1/2$ corresponding to the It\^o and Stratonovich prescriptions, respectively. Treating time as a continuous variable and evolution as the action of pullbacks induced by diffeomorphisms implies that infinitesimal evolution is generated by Lie derivatives. The structure of the Lie derivative then uniquely fixes the ordering of the position and momentum operators to the (bi-graded) Weyl symmetrization rule, corresponding to $\alpha=1/2$. Thus, the Stratonovich interpretation is the unique choice consistent with continuous-time dynamics.

\color{black}

\subsection{Eigensystem}
\label{Sec:Eigensystem}
The SEO has the following properties \cite{Ensslin_2016,Ovchinnikov_2018} (see Fig.\ref{Fig_1}a). It is a real operator, which means that its eigenvalues are either real or come in complex conjugates. This also means that SEO is pseudo-Hermitian \cite{Mostafazadeh_2002,Mostafazadeh_2002_1} and it has a complete bi-orthogonal eigensystem,
\begin{eqnarray}   
   & \langle \alpha| \beta\rangle = \delta_{\alpha\beta}, \; \hat 1 = \sum\nolimits_\alpha |\alpha\rangle \langle\alpha|,\; \hat H = \sum\nolimits_{\alpha} |\alpha\rangle H_\alpha \langle\alpha|,
\end{eqnarray}
where $\alpha$ runs over all the eigenstates and $H_\alpha$'s being the corresponding eigenvalues.

SEO commutes with the fermion number operator, $\hat k=\chi^i \partial/{\partial \chi^i}$, and each eigenstate has a well defined number of fermions, 
\begin{eqnarray}
    \hat k |\alpha\rangle = k_\alpha |\alpha\rangle, \; |\alpha\rangle\in\Omega^{(k_\alpha)},
\end{eqnarray}
where $\Omega^{(k)}$ is the subspace of Hilbert space spanned by differential forms of degree $k$, $\Omega=\bigoplus_{k=0}^{D}\Omega^{(k)}$.

SEO also commutes with $\hat d$. This divides eigenstates into a countable number of supersymmetric singlets and infinite number of non-supersymmetric doublets. 

The singlets, which we denote as $|\theta\rangle$, belong to the cohomology of $\hat d$, that is, de Rham cohomology \cite{Nakahara_book}. Each cohomology class provides one singlet, which follows from the completeness of the eigensystem of pseido-Hermitian SEO. The singlets are such that their bras and kets are annihilated by $\hat d$: $\hat d | \theta \rangle=0$ and $\langle \theta | \hat d = 0$. This means that expectation values of all $\hat d$-exact operators vanish on supersymmetric singlets:
\begin{eqnarray}
    \langle \theta | [\hat d, \hat x] | \theta \rangle =0, \forall \hat x.\label{Eq:VanishingOnSinglets}
\end{eqnarray}
SEO (\ref{Eq:DExactH}) is also $d$-exact so that all the singlets have exactly zero eigenvalue.

Non-supersymmetric doublets are related by $\hat d$ and therefore can be defined by a single bra-ket pair: 
\begin{eqnarray}
    | \gamma \rangle \text{ and } |\tilde \gamma\rangle = \hat d | \gamma \rangle, \text{ and } 
    \langle \gamma | = \langle \tilde \gamma |\hat d  \text{ and } \langle \tilde \gamma |,\label{NonSUSY_states}
\end{eqnarray}
such that $\langle \gamma|\zeta\rangle = \langle\tilde \gamma|\tilde \zeta\rangle = \langle \tilde \gamma| \hat d |\zeta\rangle = \delta_{\gamma\zeta}$. 

Using the above classification of the eigenstates, the resolution of unity and $\hat H$ take the following forms,
\begin{eqnarray}   
   & \hat 1 = \sum\nolimits_\theta |\theta\rangle \langle\theta| + \left[\hat d, \sum\nolimits_{\gamma} |\gamma\rangle \langle\tilde \gamma|\right],
\end{eqnarray}
and
\begin{eqnarray}
   & \hat H = [\hat d, \hat {\bar d}' ], \;\;\hat {\bar d}' = \sum\nolimits_{\gamma} |\gamma\rangle H_\gamma \langle\tilde \gamma|.\label{Eq:resolutionOfH}
\end{eqnarray}
\color{myblue}
\subsection{Topological supersymmetry}
\label{Sec:TS}

The commutativity of the exterior derivative with the SEO suggests that this operator is a symmetry of the model. The mathematical reason behind this symmetry is the "naturality" of the exterior derivative, i.e., the property that $\hat d$ commutes with the pullbacks of all diffeomorphisms and, consequently, with the noise-averaged pullback, which is essentially what the SEO is. This symmetry represents the fact that diffeomorphisms do not tear the fabric of phase space. For any configuration of the noise, nearby points remain nearby. In other words, this symmetry is the algebraic representation of the preservation of the topology of phase space by continuous-time dynamics. This is yet another reason why TS is a suitable identifier for this symmetry.

Viewing TS from this perspective suggests that its spontaneous breakdown implies that nearby points may no longer remain nearby in the long-wavelength limit. This is nothing other than the classical description of the butterfly effect, which indicates that spontaneous TS breaking must be associated with chaotic dynamics -- a fact that will be established rigorously in Sec.~\ref{Sec:SUSYbreaking}. In contrast to the traditional trajectory-based picture of the butterfly effect, the spontaneous-TS-breaking picture works equally well in the presence of noise.

\color{black}
\subsection{$GL(1|1)$ symmetry}
\label{Sec:SUSY}
In some classes of SDEs including Langevin SDEs \cite{Parisi_Sourlas_1979,Parisi_Sourlas_1982}, classical mechanics \cite{Gozzi_1993}, all deterministic continuous-time dynamical systems where $\hat {\bar d}=\hat \imath_f$, and the example model in Appendix \ref{Sec:Example_Model}, operator $\hat {\bar d}$ from Eq.(\ref{Eq:DExactH}) is nilpotent and commutative with $\hat H$. In these cases, $\hat{\bar d}$ can be used as the second supersymmetry operator. 

For a general form SDE, however, $\hat{\bar d}$ is not nilpotent. Nevertheless, the second supersymmetry does exist in the form of $\hat {\bar d}'$ in Eq.(\ref{Eq:resolutionOfH}). Unlike $\hat {\bar d}$, $\hat{\bar d}'$ is always nilpotent and commutative with the SEO, 
\begin{eqnarray}
    \hat{\bar d}'{}^2=0,\; [\hat H, \hat{\bar d}']=0.
\end{eqnarray}
Finding its explicit expression in terms of $\hat \Phi$ may be a complex task~\cite{Sethi_Weiderpass_2026}. This operator, however, is always available in terms of the eigensystem of SEO in Eq.(\ref{Eq:resolutionOfH}).

The symmetry operators discussed so far are the two supersymmtries, $\hat d, \hat {\bar d}'$, the ghost number operator, $\hat k$, and $\hat H$ itself having the meaning of translations in time. These operators constitute the $\mathfrak{gl}(1,1)$ superalgebra \cite{Schomerus_Saleur_2006} with non-vanishing bi-graded commutators being,
\begin{eqnarray}
    [\hat k, \hat d] = \hat d, [\hat k, \hat {\bar d}'] = -\hat {\bar d}', [\hat d, \hat {\bar d}'] = \hat H.\label{Eq:TheAlgebraOriginal}
\end{eqnarray}

The existence of the second supercharge makes it possible to express the SEO as the square of a supercharge with mixed fermion degree, in the spirit of supersymmetric theories in particle physics. This construction is traditionally used to prove that the ground-state energy must be positive when supersymmetry is spontaneously broken in models with Hermitian evolution operators. As will be discussed in Sec.~\ref{Sec:SUSYbreaking}, however, the spontaneous breakdown of TS in STS occurs when the (real part of the) ground-state eigenvalue is instead negative -- a situation with no analogue in particle physics.

It must also be emphasized that, although $\hat d$ is part of a larger supersymmetry algebra, it is important in its own right. The exterior derivative alone is sufficient to partition the Hilbert space into supersymmetric singlets and non-supersymmetric doublets. This is precisely what happens in de Rham cohomology, which relies solely on $\hat d$ and does not require help from, say, its Hodge dual, often invoked in discussions of supersymmetry breaking in models with Hermitian evolution operators.

This observation is closely related to the dual character of STS, which may be viewed both as a pseudo-unitary supersymmetric theory and as a cohomological TFT (see Sec.~\ref{Sec:TFT}). In this respect, the situation bears a certain resemblance to the notion of cohomological reduction~\cite{Candu_2010}. There, a nilpotent operator is singled out from an algebra of a larger (target space) supersymmetry (which can be spontaneously broken) and promoted to the role of a TS/BRST to extract a topological sector from the theory. In STS, the situation is reversed: the TS is central focus from the outset, and the algebra of a larger supersymmetry is a complementary structure that will be useful in construction of ETF in Sec.\ref{Sec:EFT}.
\color{black}

\review{
\subsection{$GL(1|1)$ as internal symmetry}
\label{Sec:GL(1_1)_EFT}

For the upcoming discussion in Sec.\ref{Sec:EFT}, another set of operators is more convenient than Eq. (\ref{Eq:TheAlgebraOriginal}). Namely,
\begin{eqnarray}
   &\hat Q \equiv \hat d = \sum\nolimits_{\gamma} |\tilde \gamma\rangle \langle\gamma|,\; \hat Q^T =  \sum\nolimits_{\gamma} |\gamma\rangle \langle\tilde\gamma|,\nonumber \\
   &\hat K = \sum\nolimits_{\gamma} \left( |\tilde\gamma\rangle \langle\tilde\gamma| - |\gamma\rangle \langle\gamma|\right)/2,\nonumber\\
   &\hat E =\sum\nolimits_{\gamma} (|\tilde\gamma\rangle \langle\tilde\gamma| + |\gamma\rangle \langle\gamma|) = \hat 1 - \sum_\theta|\theta\rangle\langle\theta|.
\end{eqnarray}
Here, $\hat E$ is the unity operator of and/or the projection operator on the Hilbert space spanned by non-supersymmetric eigenstates only, and we also switched notations to capitals which is in spirit of theories of high-dimensional models that will be addressed in Sec.\ref{Sec:EFT}.

These operators is also a symmetry of the model because all of them commute with $\hat H$. They also form $\mathfrak{gl}(1,1)$,
\begin{eqnarray}
    [\hat K, \hat Q] = \hat Q,[\hat K, \hat Q^T] = -\hat Q^T, [\hat Q, \hat Q^T] = \hat E.\label{Eq:TheAlgebra}
\end{eqnarray}
As will be discussed in more detail in Sec.\ref{Sec:EFT}, EFTs are theories describing slow dynamics on the Hilbert space comprised of the ground state and the states close to it, all of which are non-supersymmetric. In other words, under the condition of spontaneously broken TS, all eigenstates in the Hilbert space of the EFT are non-supersymmetric. Therefore, in the context of EFTs, the supersymmetric states can be neglected and $\hat E$ can be substituted by the ordinary unity operator,  $\hat E \stackrel{\text{EFT}}{\longrightarrow} \hat 1$.

As compared to Eq.(\ref{Eq:TheAlgebraOriginal}), Eq.(\ref{Eq:TheAlgebra}) can be identified as an "internal" symmetry, because $\hat E$ no longer represents translations in time as does its counterpart $\hat H$ in Eqs.(\ref{Eq:TheAlgebraOriginal}). It represents the freedom in choosing the bras and kets of the eigenstates of a pseudo-Hermitian evolution operator. Namely, we can perform the following transformation, $\langle \gamma| \to \langle \gamma| e^{l},\; |\gamma\rangle\to e^{-l}|\gamma\rangle$, on all the eigenstates and arrive at the same evolution operator. Operator $\hat E$ is the infinitesimal operator of this transformation. 

This has an analogue in quantum mechanics, which is invariant of the phase transformation of wavefunctions $\langle\psi|\to\langle\psi|e^{-i\varphi}, |\psi\rangle \to e^{i\varphi}|\psi\rangle$. In second quantization approach, this symmetry turns into the global internal $U(1)$ which represents the conservation of the global number of particles in higher-dimensional models. By the same token, EFTs in higher-dimensional models with broken TS must have a global internal $GL(1|1)$ symmetry.
}

\section{Weighted Traces of SEO}
\label{Sec:Weighted_Traces}
\subsection{Parisi-Sourlas functional and Witten index}
\label{Sec:ParisiSourlasConstruction}
The Parisi–Sourlas proposal is to consider the following functional
\begin{eqnarray}
    Z_\text{PS} =  \iint_\text{PBC} \delta(\dot \varphi -F) \text{Det}\frac{\delta(\dot \varphi -F)}{\delta\varphi}D\varphi P(\xi)D\xi. \label{Eq:PSconstruction}
\end{eqnarray}
With the intention of integrating out the noise variables at a later stage, Eq.(\ref{Eq:PSconstruction}) can be interpreted as a change of variables in the partition function of the noise (\ref{Eq:PartitionFunctionNoise}) from $\xi$ to $\varphi$. The periodic boundary conditions (P.B.C.) used here is the only choice that avoids introducing additional external parameters into Eq.(\ref{Eq:PSconstruction}).

Using the notation for averaging over the noise in Eq.(\ref{Eq:definitionOfNoiseAveraging}), Eq.(\ref{Eq:PSconstruction}) can be rewritten as,
\begin{eqnarray}
    Z_\text{PS} = \overline{\mathcal{I}(\xi)}, \label{Eq:ParisiSourlasFunctional}
\end{eqnarray}
where
\begin{eqnarray}
    \mathcal{I}(\xi) = \iint_\text{PBC} \delta(\dot \varphi -F) \text{Det}\frac{\delta(\dot \varphi -F)}{\delta\varphi}D\varphi.\label{Eq:Index_0}
\end{eqnarray}
The later quantity is a topological invariant, independent of $\xi$. This becomes transparent from the equivalent expression
\begin{eqnarray}
    \mathcal{I}(\xi) = \sum_{\varphi_i\in\text{closed solutions}} \text{sign Det} \frac{\delta(\dot \varphi -F)}{\delta\varphi}(\varphi_i),\label{Eq:Index}
\end{eqnarray}
which makes explicit that $\mathcal{I}(\xi)$ is the index of "vector field" $\dot\varphi-F$ defined on the space of closed paths in $X$. The index is the sum of the signs of the Jacobians at the zeros of the field. The zeros correspond to closed solutions of the SDE, $\dot\varphi=F, \varphi(0)=\varphi(T)$. This is a setting to which the Poincare–Hopf theorem applies saying that the index is a topological invariant: as the noise configuration varies, these solutions appear and disappear in pairs with opposite signs of their Jacobians leaving the index unchanged.

As it will be seen in Eq.(\ref{Eq:ParisiSourlasFunctional_3}) below, $\mathcal{I}$ is the Euler characteristic of $X$, which we denote as $\mathcal{E}(X)$ \footnote{From dynamical systems theory point of view, this index can also be identified as the Lefschetz number~\cite{Ovchinnikov_2016}.}. Therefore,
\begin{eqnarray}
    Z_\text{PS} = \mathcal{E}(X) Z_N.\label{Eq:ParisiSourlasFunctionalIndex}
\end{eqnarray}
\review{Although we use a normalized distribution of noise configurations (\ref{Eq:PartitionFunctionNoise}) such that $Z_\text{N}=1$, we keep this factor explicit in Eq.~(\ref{Eq:ParisiSourlasFunctionalIndex}) to emphasize -- anticipating the discussion in Sec.~\ref{Sec:Resolution} -- that the Parisi–Sourlas functional does not represent the partition function of the dynamical system, but rather that of the noise.}

From the perspective of QFT, Eq.(\ref{Eq:PSconstruction}) can be looked upon as a BRST gauge-fixing procedure (see, e.g., Refs.\cite{Peskin_Schroeder_1995_book,Henneaux_Teitelboim_1992_book}). To see this, one can exponentiate the delta-functional and the determinant in Eq.(\ref{Eq:Index_0}) with the help of the fields introduced in Eq.(\ref{Eq:SEO_operator}), and arrive at
\begin{eqnarray}
    \mathcal{I}(\xi) = \iint_\text{PBC} e^{\{Q,\int_o^T d\tau i\bar\chi(\dot\varphi-F)\}}D\Phi,\label{Eq:MAthaiQuillen}
\end{eqnarray}
with $Q$ defined in Eq. (\ref{Eq:SUSY}). The PBC here also apply to the ghosts, since reproducing the corresponding determinant requires them to obey the same boundary conditions as $\varphi$. 

\color{myblue} 
Eq.(\ref{Eq:MAthaiQuillen}) may be viewed as a trivial theory with vanishing action, in which the path integral over closed paths is gauge-fixed to the solutions of the SDE and $Q$ emerges as the corresponding BRST symmetry. As a result, the entire action is $Q$-exact. This property, which survives averaging over the noise in Eq(\ref{Eq:ParisiSourlasFunctional}), is a definitive characteristic of cohomological TFTs~\cite{TFT_BOOK}. 

This, in particular, means that the supersymmetry can also be regarded as a BRST symmetry. However, of the two identifications, TS appears to be preferable for the following reasons. 

First, the gauge-fixing picture does not naturally extend to the dynamical-systems interpretation of the SEO discussed in Sec.~\ref{Sec:SEO}, where this symmetry has a more fundamental mathematical meaning as the preservation of the topology of phase space by continuous-time dynamics, as discussed in Sec.~\ref{Sec:TS}. Second, the gauge-fixing perspective invoked by the term BRST naturally encourages one to regard the Parisi–Sourlas functional as the partition function. This interpretation is not correct, as will be discussed in Sec.~\ref{Sec:Resolution}, and has been a recurring source of confusion in the literature. For these reasons, we refer to this supersymmetry mostly as TS.
\color{black}

The integral over the noise in Eqs.(\ref{Eq:PSconstruction}) and (\ref{Eq:ParisiSourlasFunctional}) is Gaussian and can be easily performed leading to 
\begin{eqnarray}
    Z_\text{PS} = \iint_\text{PBC} e^{S(\Phi)}D\Phi, \label{Eq:ParisiSourlasFunctional_1}
\end{eqnarray}
with action $S$ introduced in Eq.(\ref{Eq:Action}). 

We can now switch over to the operator representation of evolution, 
\begin{eqnarray}
  Z_\text{PS} &=& \iint_\text{PBC} e^{\int_0^T d\tau\left(iB\dot\varphi + i\dot\chi\bar\chi  - H(\Phi)\right) }D\Phi \nonumber \\
  &=& \text{Tr }(-1)^{\hat k}e^{-T\hat H}.\label{Eq:PS_WittenIndex}    
\end{eqnarray}
Here, a more detailed expression of the action is given to highlight the details of the transition from the path-integral representation in the first line to the operator representation in the second. Namely, by standard arguments, the terms $iB\dot\varphi$ and $i\dot\chi\bar\chi$ indicate that momenta fields, $B$ and $\bar\chi$, correspond to momenta operators $-i\partial/\partial\varphi$ and $-i\partial/\partial\chi$ in the operator representation, while $H(\Phi)=\{Q, \bar d \} \to [\hat d, \hat{\bar d} ]=\hat H$ because $Q$ and $\bar d$ are the path-integral versions of $\hat d$ and $\hat {\bar d}$, respectively.

In  Eq.(\ref{Eq:PS_WittenIndex}), the alternating sign operator appears because PBCs are unnatural for fermionic fields. As a result, the Parisi-Sourlas functional is a path-integral representation of the Witten index, $W$ (see, e.g., Ref.\cite{MirrorSymmetry}). 

Only supersymmetric singlets contribute to $W$, while contributions from nonsupersymmetric doublets cancel out. Since singlets are in one-to-one correspondence with the de Rham cohomology classes,
\begin{eqnarray}
  Z_\text{PS} = W = \#\theta_\text{even}-\#\theta_\text{odd}=\mathcal{E}(X), \label{Eq:ParisiSourlasFunctional_3}    
\end{eqnarray}
where $\mathcal{E}(X)$ is the Euler character introduced earlier.

Moreover, the Parisi–Sourlas functional is insensitive to external perturbations of the model. This follows from the fact that, in the transition from Eq.~(\ref{Eq:PSconstruction}) to Eq.~(\ref{Eq:ParisiSourlasFunctionalIndex}), no assumption of time-translation invariance is invoked. Consequently, all derivations remain valid even when the deterministic flow $f$ is allowed to be time-dependent, as if the system were subject to external perturbation.

This property can be readily demonstrated in the operator representation. Indeed, consider the perturbed model,
\begin{eqnarray}
&&f\to f + \zeta_a \text{v}^a,\; Z_\text{PS}\to Z_\text{PS}(\zeta),\nonumber \\ 
&&Z_\text{PS}(\zeta) = \iint_\text{PBC} e^{\int_0^T d\tau \zeta_a(\tau)\{Q, v^a(\tau)\}  + S}D\Phi,
\end{eqnarray}
where $v^a(\tau) = i\bar\chi (\tau)\text{v}^a(\varphi(\tau))$ and $\zeta$ are the external probing fields. 

The dependence of $Z_\text{PS}$ on $\zeta$ can be revealed by the following functional derivatives,
\begin{eqnarray}
&&\left.\frac{\delta^l Z_\text{PS}(\zeta)}{\delta \zeta_{a_1}(t_1)...\zeta_{a_l}(t_l) }\right|_{\zeta=0} =\nonumber \\
&&= \iint_\text{PBC} e^{S} \{Q,v^{a_1}(t_1)\}...\{Q,v^{a_l}(t_l)\} D\Phi \nonumber \\
&&=\sum\nolimits_\alpha (-1)^{k_\alpha} \langle \alpha | [\hat d, \hat A]| \alpha \rangle = ... \label{Eq:partialDerivative}
\end{eqnarray}
where
\begin{eqnarray}
\hat A = e^{-T\hat H} \mathcal{T}\hat v^{a_1}(t_1) ... [\hat d, \hat v^{a_l}(t_l)],
\end{eqnarray}
with $\mathcal{T}$ being the operator of chronological ordering. We also used that $\{Q, v^a(t)\} \to [\hat d, \hat v^a(t)]$ upon switching over to the operator representation, with 
\begin{eqnarray}
    \hat v(t) = e^{t\hat H}\hat v e^{-t\hat H},\label{Eq:HeisenbergRepresentation}
\end{eqnarray} 
being the time-dependent operators in the Heisenberg representation, and that a product of any number of $\hat d$-exact operators is also a $\hat d$-exact operator $[\hat d, \hat v^{a_1}(t_1)] ... [\hat d, \hat v^{a_k}(t_k)] = [\hat d, \hat v^{a_1}(t_1) ... [\hat d, \hat v^{a_k}(t_k)]]$.

Utilizing the notations from Sec.\ref{Sec:Eigensystem} for singlets and doublets, the r.h.s. of Eq.(\ref{Eq:partialDerivative}):
\begin{eqnarray}
...&=&\sum\nolimits_\gamma (-1)^{k_\gamma} \langle \gamma | [\hat d, \hat A]| \gamma \rangle  
+\sum\nolimits_\gamma (-1)^{k_\gamma+1} \langle \tilde\gamma | [\hat d, \hat A]| \tilde\gamma \rangle  \nonumber \\ 
&&+ \sum\nolimits_{\theta} (-1)^{k_\theta} \langle \theta | [\hat d, \hat A] |\theta \rangle \nonumber\\ 
&=&\sum\nolimits_\gamma (-1)^{k_\gamma} \langle \tilde \gamma | [\hat d, [\hat d, \hat A]]| \gamma \rangle \nonumber\\ 
&&+ \sum\nolimits_{\theta} (-1)^{k_\theta} \langle \theta | [\hat d, \hat A] |\theta \rangle =0,
\label{Witten_index_Invariance}
\end{eqnarray}
where we used Eq.(\ref{Eq:VanishingOnSinglets}) and $[\hat d,[\hat d,\hat x]]=0, \forall \hat x$ which follows from nilpotency of the exterior derivative.

If the Parisi–Sourlas functional is interpreted as the partition function of the SDE, which is common in the Literature, one is led to conclude that the model does not have excited states, since 
$Z_\text{PS}$ is independent of both external perturbations and the duration of evolution. This conclusion is consistent with the logic of BRST quantization, of which the Parisi–Sourlas construction is essentially an instance. In that framework, all nonsupersymmetric states are regarded as unphysical, or “not real” states -- the very reason why the associated fermions are referred to as ghosts in the first place. 

This viewpoint, however, is clearly unsatisfactory: SDEs do possess excitations, and nonzero-eigenvalue nonsupersymmetric states play an important role, for instance, in the out-of-equilibrium Fokker–Planck evolution of the probability distribution toward the supersymmetric steady-state distribution (see below). In other words, viewing Parisi-Sourlas functional as a partition function of the SDE leads to the apparent contradiction.

\begin{table*}[t]\color{myblue}
\label{tab:Objects}
\caption{Stochastic evolution operator and related objects.}
\begin{ruledtabular}
\begin{tabular}{c|c|c|c}
SEO-related object & Path-integral rep. & Operator rep. & Value/Limit \\
\hline
\makecell[l]{-- Stochastic evolution operator, $\hat M_{T0}$;\\ Eqs.(\ref{Eq:SEO_operators}),(\ref{Eq:SEO_operator}), and (\ref{Eq:localized_states}) }& $\left.\iint e^{S(\Phi)}D\Phi\right|_{\begin{smallmatrix} \varphi(0),\chi(0)=\varphi_i,\chi_i\\ \varphi(T),\chi(T)=\varphi_f,\chi_f \end{smallmatrix} }$  & $\langle \varphi_f\chi_f|e^{-T\hat H}|\varphi_i\chi_i\rangle $ &  \textemdash  \\ 
\makecell[l]{-- Fokker Planck evolution operator, $\hat E_{T0}$;\\ Eqs.(\ref{Eq:FokkerPlanck}), (\ref{Eq:NoSUSY_loc_states}), (\ref{Eq:FokkerPlanckFinateTime})  }& $\left.\iint e^{S(\Phi)|_{\chi,\bar\chi\to0}}D\varphi D B\right|_{\begin{smallmatrix} \varphi(0)=\varphi_i\\ \varphi(T)=\varphi_f\end{smallmatrix} }$  & $\langle \varphi_f|e^{-T\hat H_\text{FP}}|\varphi_i\rangle $ &  \textemdash  \\
\makecell[l]{-- Parisi-Sourlas functional, $Z_\text{PS}$, or \\ Witten index, $W$; Eqs.(\ref{Eq:PS_WittenIndex}),(\ref{Eq:ParisiSourlasFunctional_3}) } & $\left.\iint  e^{S(\Phi)}D\Phi\right|_{\begin{smallmatrix} \varphi(0)=\varphi(T)\\ \chi(0)=\chi(T)\end{smallmatrix} }$  & $\text{Tr }(-1)^{\hat k}e^{-T \hat H}$ & $\mathcal{E}(X) Z_\text{N}$\\
\makecell[l]{-- Modified Parisi-Sourlas functional, $Z^\#_\text{PS}$;\\ Eq.(\ref{Eq:ModifiedTheory})} & $\left.\iint e^{S(\Phi) + \Xi(\varphi)}D\Phi\right|_{\begin{smallmatrix} \varphi(0)=\varphi(T) \\  \chi(0)=\chi(T)\end{smallmatrix} }$  & \textemdash & $\#\{\begin{smallmatrix} \text{closed} \\ \text{solutions} \end{smallmatrix} \}$\\ 
\makecell[l]{-- Partition function, $Z_\text{DS}$; Eqs.(\ref{Eq:Z_DS}),(\ref{Eq:Growth_pressure})}& $\left.\iint e^{S(\Phi)}D\Phi\right|_{\begin{smallmatrix} \varphi(0)=\varphi(T) \\ \chi(0)=-\chi(T) \end{smallmatrix} }$  & $\text{Tr }e^{-T\hat H} $ & $\stackrel{T\to\infty}{\propto} e^{\Delta T}$\\ 
\makecell[l]{-- Partition function of traditional \\ theory of SDEs, $Z_0$; Eqs.(\ref{Eq:Z_0}),(\ref{Eq:Z_0limit})} & $\left.\iint e^{S(\Phi)|_{\chi,\bar\chi\to0}} D\varphi D B \right|_{\varphi(0)=\varphi(T)}$  & $\left.\text{Tr}\right|_{\Omega^{(D)}}e^{-T\hat H} $ & $\stackrel{T\to\infty}{\to} 1$ \\ 
\makecell[l]{-- Matrix element of evolution \\ of initial condition, $Z_\text{I}$; Eq.(\ref{Eq:InitialCZ_I})} & $\left.\iint e^{S(\Phi)}D\Phi\right|_{\varphi(0)=\varphi_0} $  & $\langle 0| e^{-T\hat H}|\varphi_0 \bar 1\rangle $ & $1$ \\ 
\makecell[l]{-- Matrix element on instanton, $O_{ab}$; \\Eqs.(\ref{Eq:Instanton_Pathintegral}), (\ref{Eq:Instanon_integral_modulii}) }& $\left.\iint O(\varphi\chi) e^{S(\Phi)}D\Phi \right|_{\varphi(\pm\infty)=\varphi_{b,a}}$  & $ \langle b| \hat O |a\rangle$ & $ \sim\int_{I_{ab}} O$\\ 
\makecell[l]{-- Generating functional, $e^{ G(\zeta) }$; Eq.(\ref{Eq:GF})} & $e^{TH_g} \iint \bar\Psi_g(T)  e^{\int d\tau\zeta_a(\tau)O^a(\Phi(\tau))+S(\Phi)} \Psi_g(0)D\Phi$  & $ \langle g| \mathcal{T} e^{\int d\tau \zeta_a(\tau)\hat O^a(\tau)} |g\rangle $ &  $\stackrel{\zeta\to0}{\to} 1$ 
\end{tabular}
\end{ruledtabular}
\end{table*}

\subsection{Modified Parisi-Sourlas model}

A common way to circumvent this issue is to regard the Parisi–Sourlas construction as an approximation~\cite{Nakayama_2025,Nicolis_2026} to the other partition function obtained from Eq.~(\ref{Eq:PSconstruction}) by replacing $\mathrm{Det} \to |\mathrm{Det}|$, or, at the level of the action, $S \to S + \Xi$, where $\Xi = \log \mathrm{sign}\,\mathrm{Det}$. The modified partition function (c.f. Eqs.(\ref{Eq:ParisiSourlasFunctional}) and (\ref{Eq:Index})) can then be given as
\begin{eqnarray}
Z^\#_{\mathrm{PS}} = \overline{\mathcal{I}^\#(\xi)}, \qquad \mathcal{I}^\#(\xi) = \sum_{\varphi_i\in\text{closed solutions}} 1.\label{Eq:ModifiedTheory}
\end{eqnarray}
It admits a clear interpretation as the number of closed solutions of the SDE, averaged over the noise. From this perspective, the modified model in Eq.(\ref{Eq:ModifiedTheory}) appears reasonable.

Nevertheless, there are two important points to note regarding Eq.~(\ref{Eq:ModifiedTheory}). First, due to the presence of $\Theta$, the model is no longer supersymmetric and can no longer be identified with a Parisi–Sourlas construction. The second point is a substantial mathematical drawback: $\Theta$ is highly nonlocal in time. As a consequence, there is no associated local-in-time evolution operator of the Fokker–Planck type. Moreover, even if such an operator could be defined, it would still differ from the SEO introduced in Eq.~(\ref{Eq:SEO_operators}). From a mathematical standpoint, the noise-averaged SDE-defined pullback in Eq.~(\ref{Eq:SEO_operators}) is, however, unique and admits no alternatives. In this sense, Eq.~(\ref{Eq:ModifiedTheory}) is unlikely to provide a formulation of stochastic dynamics that is fully satisfactory from a mathematical perspective.

It should be further emphasized that the concept of a local-in-time evolution operator is foundational, particularly in quantum physics. It embodies the idea that the wavefunction at any given instant contains sufficient information about the system's past to determine its subsequent evolution uniquely. Within STS, this principle is realized in connection with the infinitely long dynamical memory associated with chaos -- the butterfly effect -- which is encoded in the fermionic sector of the wavefunction. This observation points out once again that the traditional theory of stochastic dynamics, lacking fermions, does not possess the structure required to encode the butterfly effect in its wavefunction and therefore cannot provide a complete description of dynamical chaos.

The final remark we would like to make in connection with the modification of the PS model discussed in this section concerns the following expression -- or, more precisely, one of its variants -- which is sometimes encountered in the literature on Langevin SDEs:
\begin{eqnarray}
    \iint_\text{PBC} \delta(\dot\varphi -F_L) \left|Det \frac{\delta (\dot\varphi -F_L)} {\delta \varphi}\right| D\varphi = 1,\label{Eq:wrong_identity}
\end{eqnarray}
where $\varphi\in \mathbb{R}^D$ and the flow vector field is given by the gradient of a function $U$, which we can assume to be Morse-type, together with additive noise, $F_L=-\partial U + \sqrt{2\Theta}\xi$.

The $0+0$-dimensional counterpart of the l.h.s. of Eq.(\ref{Eq:wrong_identity}) evaluates to the number of critical points of $U$,
\begin{eqnarray}
\int \delta(\partial U) \left|Det \partial\partial U \right| D\varphi = \# \{\varphi_\alpha| \partial U(\varphi_\alpha) = 0\}.
\end{eqnarray}
For nonlinear potentials, this number can be arbitrarily large.

The same result holds for Eq.~(\ref{Eq:wrong_identity}) in the deterministic limit. Indeed, in this limit each critical point of $\partial U$ gives rise to a closed trajectory, and each such trajectory contributes unity to the functional integral on the l.h.s. of Eq.(\ref{Eq:wrong_identity}).

Consequently, Eq.~(\ref{Eq:wrong_identity}) is not valid in general. It holds only in the special case in which $U$ possesses a single critical point. A Langevin SDE with a Gaussian potential provides a representative example of this class. The other members of the class are equivalent to it up to perturbative corrections.

Moreover, within this restricted class of models, the absolute value around the functional determinant can be omitted, since the determinant is always positive. Equation~(\ref{Eq:wrong_identity}) then reduces to the index in Eq.~(\ref{Eq:MAthaiQuillen}) as applied to this class of models.

We therefore conclude that Eq.~(\ref{Eq:wrong_identity}), when interpreted as a general identity, is incorrect; in the restricted class of simple models for which it does hold, it provides no information beyond that contained in Parisi-Sourlas approach.

\subsection{Parisi-Sourlas functional and the partition function of the noise}
\label{Sec:Resolution}

The resolution of the contradiction outlined in the end of Sec.~\ref{Sec:ParisiSourlasConstruction} rests on the following observation. As discussed in Sec.~\ref{Sec:Stochastics_Statistics}, the DS and its noise are two distinct dynamical systems, each with its own partition function. At the same time, Eq.~(\ref{Eq:SDE}) directly defines the partition function of the noise, but it does not directly define the partition function of the DS. Rather, it defines open-ended dynamics, and the SEO in Eq.~(\ref{Eq:SEO_operators}), which corresponds to open boundary conditions, provides a unique representation of this dynamics on $\Omega$. It is this SEO that constitutes the fundamental object defining the model. All other quantities associated with the SDE must be derived from it. These quantities may have different meanings, and one of them may serve as a representative of the partition function of the noise.

If such a representative exists, it must be independent of the duration of the time evolution, just like $Z_N$ itself, and it must be insensitive to perturbations, since the noise carries no information about the dynamics of the DS, whether perturbed or not, as there is no backaction from the DS to the noise. 

These are precisely the properties of the Parisi--Sourlas functional, from which we conclude that $Z_\text{PS}$ is a representative of the partition function of the noise, not of the DS. This interpretation is further supported by Eq.~(\ref{Eq:ParisiSourlasFunctionalIndex}), which shows that $Z_\text{PS}$ coincides with $Z_N$ up to a topological factor, as well as by the fact that this construction amounts to a mere change of variables in $Z_N$.

\subsection{Partition Function of DS}
\label{Sec:PartitionFunction}
Now it becomes clear how to construct the partition function of the DS. The source of the topological character of the Witten index is in the alternating sign operator in Eq.(\ref{Eq:PS_WittenIndex}). We just have to remove it from the trace. In the path-integral representation, this operation is equivalent to twisting the boundary conditions for the ghosts from periodic (PBC) to anti-periodic (APBC),
\begin{eqnarray}
Z_\text{DS}(T) = \iint_{\text{APBC}} e^{S(\Phi)} D\Phi = \mathrm{Tr}\, e^{-T\hat H}. \label{Eq:Z_DS}
\end{eqnarray}
Unlike the modified model in Eq.~(\ref{Eq:ModifiedTheory}), the twist does not alter the dynamics which is still governed by the SEO. Moreover, Eq.(\ref{Eq:Z_DS}) equals the number of closed solutions, but only in the limit of infinitely long evolution~\cite{Ovchinnikov_2016},
\begin{eqnarray}
\left. Z_\text{DS}(T) \right|_{T\to\infty} \propto e^{\Delta T}, \label{Eq:Growth_pressure}
\end{eqnarray}
where $\Delta = -\min_\alpha \mathrm{Re}\, H_\alpha \ge 0$ is (up to a sign) the lower bound of the SEO spectrum. In DS theory, this quantity is referred to as “pressure”~\cite{Ruelle_2002}, a generalization of the concept of dynamical entropy in deterministic DS (see, e.g., Refs.\cite{Hasselblatt_Katok_2002_book,Pesin_1977,Arnold_1988,Baxendale_1986}) to random dynamics. \color{myblue} A positive value of $\Delta$ signals exponential growth in the number of closed solutions in the long-time limit. This is a hallmark of dynamical chaos in dynamical systems theory.

One way to see that situations with $\Delta>0$ are realizable is through the connection between STS and kinematic dynamo theory, where such spectra have long been known, with both real and complex eigenvalues of the fastest-growing states (see Fig.\ref{Fig_1}a)~\cite{Ensslin_2016,Ovchinnikov_2018}. Also, $\Delta$ cannot be negative in our case because supersymmetric states always exist (see below). Therefore, $\Delta$ can either equal zero or be positive, which corresponds to chaos in the sense of random DSs~\cite{Ruelle_2002}.


\subsection{Chaos as a new type of spontaneous supersymmetry breaking}
\label{Sec:SUSYbreaking}

When $\Delta$ is positive -- or, equivalently, when the model is chaotic in the sense of random dynamical systems theory -- either one or two eigenvalues exist with real part $-\Delta<0$ (see Fig.\ref{Fig_1}a). The contribution from the corresponding doubly-degenerate states dominate the partition function (\ref{Eq:Z_DS}) in the long-time limit. The ground state of the model -- the one through which the generating functional in Sec.\ref{Sec:Generating_Functional} will be defined -- must be chosen from this set of states so that the response of the model is stable with respect to excitations to all other states. Therefore, the ground state has a nonzero eigenvalue and is consequently non-supersymmetric, as follows from Sec.~\ref{Sec:Eigensystem}.

This situation by definition corresponds to the spontaneous breakdown of TS. The term \emph{spontaneous} refers to the fact that the symmetry in question is broken by the ground state (i.e., by the system itself), in contrast to \emph{explicit} symmetry breaking, which arises from an "external" non-symmetric term in the action, as in the Zeeman effect, where an external magnetic field breaks the $O(3)$ rotational symmetry of an atom.

An equivalent definition of spontaneous TS breaking, that can also be encountered in literature, is the existence of nonvanishing expectation values of $d$-exact operators in the ground state. In our setting, one such operator is the SEO itself (see Eq.(\ref{Eq:DExactH})),
\begin{eqnarray}
    &\langle g | [\hat d, \hat{\bar d}] | g \rangle  = H_g \ne 0, \text{Re }H_g < 0, \label{Eq:TS_breaking_SDE}
\end{eqnarray}
with $|g\rangle$ denoting the ground state and $H_g$ being its eigenvalue, so that TS is spontaneously broken by this definition too. In other words, dynamical chaos of random dynamical systems theory \cite{Ruelle_2002} is equivalent to the spontaneous TS breaking in STS.

This identification is more than a mere change of terminology. Viewing chaos as spontaneous TS breaking immediately suggests that a Goldstone theorem applies, implying the existence of gapless excitations. Their presence entails long-range correlations, which are naturally associated with the experimental signature of chaos -- $1/f$ noise -- as we discuss in more detail in Sec.~\ref{Sec:EFT} below.

\subsubsection{Key role of pseudo-Hermiticity}
Spontaneous supersymmetry breaking in STS is qualitatively different from the same phenomenon in its sister theories with real, non-negative spectra -- namely, the supersymmetric $1+0$ nonlinear sigma model~\cite{Witten_1982}, topological quantum mechanics~\cite{Losev_2005,Faddeev_1993,Freed_1999} and the Parisi-Sourlas theory of Langevin SDEs. In those cases, the spectrum of the evolution operator is real and non-negative, and supersymmetric states with zero eigenvalue are always the ground states of the model. Consequently, spontaneous breakdown of supersymmetry would imply that zero-eigenvalue supersymmetric states had disappeared, which would certainly be concerning, as it would suggest that the eigensystem of the evolution operator is incomplete in the sense that it fails to account for de Rham cohomology classes.

In STS, by contrast, the eigenvalues of the pseudo-Hermitian SEO are not constrained to have non-negative real parts. Negative real parts are therefore allowed, and when such states appear, TS is spontaneously broken not because supersymmetric states disappeared, but because they are no longer ground states; instead, they become excited states relative to the non-supersymmetric ground state with a negative (real part) of its eigenvalue.

Thus, spontaneous TS breaking in STS is a direct consequence of the pseudo-Hermiticity of the SEO and therefore has no counterparts in previously studied supersymmetric theories. 

Another qualitative difference is that the very existence of the symmetry-broken phase is not in question in STS: dynamical chaos and the positivity of $\Delta$ in chaotic models have already been firmly established experimentally and theoretically. 

From the perspective of high-energy physics, the novelty of this supersymmetry-breaking mechanism and its established experimental realization underscore the significance of this connection to dynamical systems theory, which may enable a fruitful exchange of ideas between the two fields.

\subsubsection{Pseudo-time-reversal symmetry breaking}
\label{Sec:etaTbreaking}

Pseudo-Hermiticity of the SEO gives rise to yet another distinct qualitative phenomenon. It occurs when the fastest-growing eigenstates possess complex eigenvalues (see Fig.~\ref{Fig_1}a). This situation can be interpreted as spontaneous breakdown of pseudo-time-reversal-, or $\eta$T-symmetry~\cite{Mostafazadeh_2002,Mostafazadeh_2002_1}, since eigenstates with complex-conjugate eigenvalues form $\eta$T-pairs. Selecting one of them as the ground state seemingly breaks this pairing symmetry. In kinematic dynamo theory, this corresponds to the rotation of galactic magnetic fields~\cite{Ensslin_2016,Ovchinnikov_2018,Bouya_2013}. By analogy, it may also correspond to a form of rotation in chaotic dynamics.

This phenomenon may have a broader physical significance. For example, it may be related to the notion of the "arrow of time."
\subsubsection{STS as TFT}
\label{Sec:TFT}
It is also instructive to examine the implications of spontaneous TS breaking for the Baulieu–Grossman identification of STS as a TFT. To begin with, it should be emphasized that there are two broad classes of TFTs -- quantum and cohomological~\cite{TFT_BOOK}. STS is in the latter category. In a cohomological TFT, only certain observables are topological invariants, rather than the entire theory being independent of geometric details. In the case of STS, such topological quantities include the Parisi–Sourlas functional and/or the Witten index.

The identification of a theory as a cohomological TFT requires the presence of several key structures. Foremost among them are a nilpotent supersymmetry, $Q$, commonly called a TS or a BRST symmetry, and a $Q$-exact action. Both ingredients are unconditionally present in STS.

In addition, one must be able to identify a set of states, $|\rho\rangle, \rho=1,2,...$, that are supersymmetric in the sense that, $\langle \rho | \{Q, x\}| \rho'\rangle =0 , \forall x$, and a set of observables, $O_a, a=1,2,...$, that are $Q$-closed but not $Q$-exact. Once such states and observables have been identified, the corresponding matrix elements, $\langle \rho |  O_{a_1} ... O_{a_k} | \rho'\rangle$, are either topological invariants or vanish identically.

There are two natural choices for the realization of TFT within STS. The first is associated with instantons (see Sec.~\ref{Sec:Instantons} below). In this case, the supersymmetric states are the perturbative (local) supersymmetric states corresponding to the critical points (or invariant sets) of the deterministic flow, while the observables are Poincar\'e duals \footnote{A Poincare dual of a submanifold, $M$, is a differential form, $p(M)\in\Omega^{(D-\text{dim} M)}$, satisfying $\int_X p(M)\wedge \omega =\int_{M}\omega$ for any $(\text{dim }M)$-form $\omega$ in $X$.} of closed submanifolds of the phase space. In the deterministic limit, the corresponding matrix elements reduce to intersection numbers on the instanton manifolds. 

This version of TFT within STS survives spontaneous TS breaking, provided the deterministic flow remains integrable and the concept of instantons is valid. Such a situation is realized in noise-induced or instantonic chaos (see Fig.~\ref{Fig_1}b and Appendix~\ref{Sec:Example_Model}), suggesting that the corresponding topological matrix elements may have useful applications.

The second natural choice of states and observables is provided by the global supersymmetric states of Eq.~(\ref{Eq:VanishingOnSinglets}) together with observables which are Poincar\'e duals of homologically nontrivial submanifolds of the phase space. The corresponding matrix elements are the structure coefficients of the de Rham cohomology ring. Unlike the instantons, this realization of TFT within STS is always present: it persists even when TS is spontaneously broken and/or the deterministic flow is nonintegrable.

When TS is unbroken, the situation is fundamentally no different from that in the unitary counterparts of STS~\cite{Witten_1982,Losev_2005,Faddeev_1993,Freed_1999}. When TS is spontaneously broken, however, the global supersymmetric states and the associated topological invariants cease to be directly related to the "sustained dynamics" (see Sec.\ref{Sec:StateToState}) taking place around the non-supersymmetric ground state -- the dynamics described by the EFT. 

In must be pointed out, however, that EFTs may still contain a nontrivial topological structure as suggested by the concept of cohomological reduction~\cite{Candu_2010} and as speculated in Sec.~\ref{Sec:EFT}.

\color{black}

\subsection{Traditional theory of SDEs}
\label{Sec:PartitionFunction}

Let us now establish the connection to the traditional theory of SDEs describing the temporal evolution of probability distributions. Within STS, probability distributions are represented, in a coordinate-free setting, as top differential forms from $\Omega^{(D)}$. The Fokker-Planck evolution operator is the SEO on $\Omega^{(D)}$,
\begin{eqnarray}
    \hat H_\text{FP} = \left. : \hat H|_{\Omega^{(D)}} : \right|_{\chi,\hat{\bar\chi}\to 0}
    = \frac{\partial}{\partial \varphi^i} \left( f^i - g_a^i \frac{\partial}{\partial \varphi^j} g_a^j \right), \label{Eq:FokkerPlanck}
\end{eqnarray}
where the colons denote normal ordering with respect to the fermionic operators, such that all $\chi$ are placed to the right of all $\hat{\bar\chi}$. After the normal ordering, the fermionic operator are dropped, and all wavefunctions can be taken to depend only on the bosonic variables, $|\psi\rangle=\int_Xd\varphi\psi(\varphi)|\varphi\rangle$, where $|\varphi\rangle$ denotes a localized state,
\begin{eqnarray}
    \langle \varphi'| \varphi\rangle = \delta^D(\varphi-\varphi').\label{Eq:NoSUSY_loc_states}
\end{eqnarray}
The finite-time Fokker-Planck evolution operator can be now expressed as~\footnote{The second equality in this equation holds only up to the path-integral ambiguity associated with It\^o-Stratonovich dilemma.},
\begin{eqnarray}
\hat E_{T0} = e^{-T\hat H_\text{FP}}=\iint_{\mathrm{open BC}} e^{S(\Phi)|_{\chi,\bar\chi\to0}}D\varphi DB,\label{Eq:FokkerPlanckFinateTime}
\end{eqnarray}
where the action in the path integral is obtained from Eq.(\ref{Eq:Action}) by setting all fermionic fields to zero: $S(\Phi)|_{\chi,\bar\chi\to0} = \int_0^T d\tau iB_i(\dot \varphi^i - (f^i - i B_j g^j_a g^i_a) )$. The corresponding partition function is
\begin{eqnarray}
    Z_0 = \iint_\text{PBC} e^{S(\Phi)|_{\chi,\bar\chi\to0}}D\varphi DB =\text{ Tr } e^{-T\hat H_\text{FP}}.\label{Eq:Z_0}
\end{eqnarray}
The Fokker-Planck evolution operator always has a zero-eigenvalue steady-state probability distribution often referred to as ergodic zero, which, in STS, is a supersymmetric state from $\Omega^{(D)}$. Returning to the coordinate-free setting, the ergodic zero can be expressed as,
\begin{eqnarray}
    |0\rangle = P_\text{ss}(\varphi)\chi^1...\chi^D \text{ and }  \langle 0| = 1.
\end{eqnarray}
Its bra is a constant function on $X$ so that $\langle 0| 0 \rangle = \int_X P_\text{ss} (\varphi)\chi^1...\chi^Dd^D\varphi d^D\chi = \int_X P_\text{ss} (\varphi)d^D\varphi =1$, which is essentially the normalization condition for $P_\text{ss}$.

The real parts of the eigenvalues of all other (nonsupersymmetric) eigenstates in $\Omega^{(D)}$ are positive. Therefore,
\begin{eqnarray}
    Z_0(T)\big|_{T\to\infty} = \langle 0| e^{-T\hat H} |0\rangle = 1. \label{Eq:Z_0limit}
\end{eqnarray}

Comparing Eqs.~(\ref{Eq:Z_0limit}) and (\ref{Eq:Growth_pressure}), one concludes that, unlike STS, the traditional theory of SDEs fails to capture the essence of chaos. The reason is straightforward: TS can be spontaneously broken only by eigenstates with nontrivial fermionic content -- an observation that may be viewed as a precursor to the Goldstone theorem and to ghost condensation in QFT. Such states are, however, absent in the traditional theory of SDEs.

\section{Physicality of ghosts}
\label{Sec:Physicality}

In QFT, the twist of the boundary conditions that leads from Eq.~(\ref{Eq:PS_WittenIndex}) to Eq.~(\ref{Eq:Z_DS}) amounts to promoting non-supersymmetric states to physical status. This notion of physical states is a subtle yet central aspect of QFT (see Sec.~3.6 of Ref.~\cite{TFT_BOOK}). However, the arguments that define physicality in that context are tailored to QFTs and do not readily carry over to other settings. For example, the spin–statistics theorem has little relevance for DSs such as financial markets, even though they too can be described by SDEs.

For this reason, the question of what should be regarded as a “physical state” must be reconsidered in the context of stochastic dynamics. At a basic level, this amounts to asking how meaningful the STS truly is in comparison to, say, the traditional theory of SDEs. It is useful to separate this question into two parts: a mathematical one and one of interpretation.

From the mathematical standpoint, the answer is clearly affirmative. The STS succeeds precisely where the traditional theory falls short: it captures the essence of chaos. This is evident, for example, from a comparison of Eqs.~(\ref{Eq:Z_0limit}) and (\ref{Eq:Growth_pressure}), which shows that STS reproduces the characteristic exponential growth associated with chaotic behavior. In fact, this is one of the reasons why random DSs theory has long employed the concept of evolution of differential forms~\cite{Ruelle_2002}.

Moreover, unlike tradition theory of SDEs, STS contains a prominent representative -- the Witten index -- of the partition function of the noise, a most fundamental object that already appears at the very level of defining the SDE itself. These features indicate that STS is not merely a formal construction, but a mathematically consistent and practically useful framework.

The question of interpretation of STS is essentially the question of the interpretation of the ghosts~\footnote{The author would like to thank Torsten En{\ss}lin and Jens Jasche for the discussion about this issue.}\cite{Ovchinnikov_2024}. A good starting point toward such interpretation is the meaning of the probability distributions in the traditional theory of SDEs. What does this distribution actually represent? Apparently, it does not describe the DS itself: in real world there is only one realization of the noise and a real DS is always deterministic. Rather, the probability distribution encodes our uncertainty about DS.

This point becomes clearer in a more concrete setting. Consider a numerical experimentalist studying a stochastic system, say, in astrophysics or finance. At the initial time, the researcher knows the initial condition of the DS, $\varphi_0$, and aims to predict its future evolution. In this language, the initial state of knowledge is represented by the probability distribution,
\begin{eqnarray}
|P(0)\rangle = \delta^D(\varphi-\varphi_0)d\varphi^1\wedge...\wedge d\varphi^D= |\varphi_0 \bar 1\rangle,  \label{Eq:P0}
\end{eqnarray}
where we recalled again that, in the coordinate-free setting, probability distributions are top differential forms. Here we introduced two new notations: the volume differential, $\bar1 = \chi^1...\chi^D$, and the notation for localized states $|\varphi\chi\rangle$ such that  
\begin{eqnarray}
    \langle\varphi'\chi'|\varphi\chi\rangle=\delta^D(\varphi-\varphi')\delta^D(\chi-\chi').\label{Eq:localized_states}
\end{eqnarray}

Each possible realization of the noise defines a unique trajectory, including its endpoint. This one-to-one correspondence between noise configurations and trajectories ensures that the integration over trajectories is well defined. Its mathematical representation is given by the SEO in Eqs.(\ref{Eq:SEO_operators}) and (\ref{Eq:SEO_operator}). In other words, the observer can numerically obtain the probability distribution for their uncertainly about DS at a future time $T$ as
\begin{eqnarray}
|P(T)\hat 1\rangle = \hat M_{T0} |\varphi_0\bar 1\rangle.
\end{eqnarray}
The SEO preserves the normalization, so that
\begin{eqnarray}
\int_X P(T) = \langle 0 | \hat M_{T0} |\varphi_0\bar 1\rangle = 1,
\end{eqnarray}
or, in the path-integral representation,
\begin{eqnarray}
Z_\text{I} = \langle 0 | \hat M_{T0} |\varphi_0\bar 1\rangle = \left.\iint e^{S(\Phi)}D\Phi\right|_{\varphi(0)=\varphi_0}=1.\label{Eq:InitialCZ_I}
\end{eqnarray}
This is one of the path-integral representations of an SDE in the literature~\cite{Sethi_Weiderpass_2026}. \review{Coupling the model to fermionic perturbations will promote this matrix element into a generating functional describing the evolution not only on the top-degree differential forms but also on other wavefunctions with nontrivial fermionic content. In chaotic models, the associated response functions will exhibit exponential growth in time, reflecting the presence of positive Lyapunov exponents~\cite{Ovchinnikov_2024}.}

Returning to the interpretation of ghosts, the probability-distribution description of a model may not be sufficient for all observers. Some may also be interested in Lyapunov exponents, which quantify sensitivity to initial conditions. Accessing this information requires propagating infinitesimal differentials alongside $\varphi$. To avoid redundancy, these differentials must be antisymmetric, i.e., Grassmann-valued. The ghosts of STS provide precisely the appropriate tool for this purpose~\cite{Graham_1988,Gozzi_1994_Lyapunov}.

From this perspective, the meaning of wavefunctions in STS becomes transparent. A state such as $|\varphi_0\chi\rangle$, represents an observer who, in addition to specifying the initial condition, $\varphi_0$, intends to track a set of differentials, $\bar\chi$~\footnote{There is a subtlety here that the differentials in the numerical experiment are dual to the differentials in the wavefunction: $\overline{\chi^{i_1}...\chi^{i_k}}=e^{i_1..i_k}{}_{i_{k+1}..i_D}\chi^{k+1}..\chi^{D}$, with $e$ being the antisymmetric tensor \cite{Ovchinnikov_2024}}. After the time evolution understood again as a numerical integration over all noise configurations and/or trajectories, the state of knowledge of the observer about the model at the future time $T$ becomes,
\begin{eqnarray}
\hat M_{T0}|\varphi_0\chi\rangle,
\end{eqnarray}
thereby encoding not only the probabilistic distribution over the original DS variables $\varphi$, but also the geometric structure of their sensitivity to initial conditions.

\section{State-to-state matrix elements}
\label{Sec:StateToState}

The state-to-state matrix element~(\ref{Eq:InitialCZ_I}) has a clear interpretation as an instance of a numerical experiment. It is a more natural representative of stochastic dynamics than the weighted traces of the SEO in Eqs.~(\ref{Eq:Z_DS}), (\ref{Eq:Z_0}), and (\ref{Eq:PS_WittenIndex}). In fact, there are at least two other types of state-to-state matrix elements that have a clear meaning from the perspective of stochastic dynamics.

The first is the (global) ground-state-to-ground-state matrix element, which represents what is often called \emph{self-sustained} dynamics. It is this object that must be employed in constructing effective field theories of chaos, as will be discussed in Secs.~\ref{Sec:Generating_Functional} and \ref{Sec:EFT}. The second is the (perturbative) vacuum-to-vacuum matrix elements known as instantons, which represent \emph{transient} dynamics and to which we now turn.

\subsection{Instantons}
\label{Sec:Instantons}

Instantons can be understood as “quanta” of strongly nonlinear, transient dynamics. Earthquakes, neuronal avalanches, and bubble-wrap pops are all familiar examples of such events in real systems. They are important within STS for two reasons. First, instantons play a central role in what may be termed noise-induced chaos, also known as the edge of chaos~\cite{Langton_1990,Mitchell_1993,Kauffman_1993,Crutchfield_2003,Crutchfield_2012} -- a special phase at the boundary of deterministic chaos in which TS is spontaneously broken not by the non-integrability of the flow, as in conventional deterministic chaos, but through the condensation of noise-induced instantons~\cite{Ovchinnikov_2024}.

Second, Witten-type TFTs -- and STS as a member of this family -- provide a convenient framework for computing certain topological invariants as matrix elements on instantons~\cite{Frenkel_2007,MirrorSymmetry,Anselmi_1997}. In fact, this aspect of TFTs is so compelling that TFT are sometimes identified as intersection theories on instantons. Therefore, any discussion of objects related to the SEO, or of the topological properties of STS, would be incomplete without a brief discussion of instantons.

With the help of already introduced notations, an instantonic matrix element can be expressed as,
\begin{eqnarray}
    O_{ab} = \left.\iint O(\varphi\chi)e^{S}D\Phi\right|_{\varphi(\pm\infty)=\varphi_{b,a}}.\label{Eq:Instanton_Pathintegral}
\end{eqnarray}
Here, $\varphi_{a,b}$ are two critical points of the deterministic part of the flow, $f(\varphi_{a,b})=0$, \review{and $O(\varphi\chi)$ is some functional of $\varphi$ and $\chi$}. For simplicity, we can focus only on Morse-like flows, in which critical points are isolated. 

If two critical points are connected by deterministic trajectories, $\dot\varphi(t)=f(\varphi(t)), \varphi(\pm\infty)=\varphi_{b,a}$, the trajectories come in families that, from the set-theoretic perspective, form a manifold, $I_{ab}$, of dimension $\mathrm{dim} I_{ab} = \mathrm{ind} a - \mathrm{ind} b$, where $\mathrm{ind}$ stands for index defined as a number of unstable directions of the flow at a critical point. It is this manifold which is called instanton, or, more accurately, intstanton modulii space.

\review{The path integral in Eq.~(\ref{Eq:Instanton_Pathintegral}) reduces to an integral over $I_{ab}$ in accordance with the localization principle in supersymmetric theories (see, e.g., Ref.~\cite{MirrorSymmetry}). In the one-loop approximation -- which is exact in the deterministic limit assumed throughout the remainder of this section -- the determinants from fluctuations in bosonic and fermionic variables cancel each other, yielding
\begin{eqnarray}
    O_{ab} = \int_{I_{ab}} O, \label{Eq:Instanon_integral_modulii}
\end{eqnarray}
where $O$ in the r.h.s. is a restriction of $O(\varphi\chi)$ to $I_{ab}$. In anticipation of this, $O$ in Eq.(\ref{Eq:Instanton_Pathintegral}) was assumed a functional of $\varphi$ and $\chi$, but not the momenta fields, $B$ and $\bar\chi$. Therefore, interpretation of $O$ as a differential form is straightforward.}

The topological character of matrix elements on instantons is most easily seen in the operator representation where Eq.(\ref{Eq:Instanton_Pathintegral}) can be expressed as
\begin{eqnarray}
    O_{ab} = \langle b| \hat O | a \rangle,\label{Eq:InstantonMatrixElement}
\end{eqnarray}
with $\langle a|$ and $|b\rangle$ being the bra and ket of the so called perturbative supersymmetric states, or vacua, associated with these critical points. These states form the Morse–Smale–Witten complex~\cite{MirrorSymmetry}. They can be defined as, 
\begin{eqnarray}
    \langle b| = p(S_b), \text{ and } |a\rangle = p(U_a),
\end{eqnarray}
where $S/U$ are the stable/unstable manifolds, respectively -- i.e., the sets of points that flow to the given critical point under $f$ in the infinite future/past. \review{Here, notation $p$ stands for Poincare dual.}

In these terms, the instanton manifold is the intersection of the corresponding stable and unstable manifolds, $I_{ab} = S_b \cap U_a$, -- an observation that leads from Eq.(\ref{Eq:InstantonMatrixElement}) to Eq.(\ref{Eq:Instanon_integral_modulii}).

Now we can consider a set of submanifolds of the phase space, $\eta_l\in X, l=1,...,n$, and their Poincare duals, $p(\eta_l)\in\Omega^{(D-\dim\eta_l)}$. \review{In the operator representation (\ref{Eq:HeisenbergRepresentation}), the duals give raise to time-dependent operators, 
\begin{eqnarray}
 \hat \eta_l(t)= e^{t\hat H} p(\eta_l) e^{-t\hat H}= p(\eta_l(t)), \label{Eq:Observables}
\end{eqnarray}
where $\eta_l(t)$ is the image of $\eta_l$ under the flow for duration $t$. Eq.(\ref{Eq:Observables}) utilizes that in the deterministic limit accepted here, SEO consists of only the Lie derivative: $\hat H \to \hat L_f$. 

It is now clear that the matrix element,
\begin{eqnarray}
    &\langle b| \hat O | a\rangle = \int_X p(S_b) \wedge p(\eta_{l_1}(t_1))..p(\eta_{l_k}(t_k)) \wedge p(U_a) \nonumber \\ &= \int_{I_{ab}} p(\eta_{l_1}(t_1))\wedge ...\wedge p(\eta_{l_k}(t_k)), 
\end{eqnarray}
with $\hat O  = \hat\eta_{l_1}(t_1)...\hat\eta_{l_k}(t_k) $, represents the intersection number of the corresponding submanifolds on $I_{ab}$.} The number is defined as a sum of the signs of the orientations of the submanifolds over all the intersection points. We also utilized that $e^{-t\hat H}|a\rangle =|a\rangle$ and $\langle b| e^{t\hat H} =\langle b|$ because $I_{ab}, S_b$, and $U_a$ are all invariant under the flow, $f$.

As we vary $t$'s, the intersection points appear and disappear in pairs with opposite orientation so that the number remains the same. In other words, it is independent of $t$'s. This is the topological invariance in its simplest setting when the basespace consists only of time.

\subsection{Generating Functional}
\label{Sec:Generating_Functional}
One of the most important concepts in QFT and statistical physics is the generating functional (GF). This object contains information on how the model responses to external perturbations. In statistical physics, the GF is associated with the partition function of a system perturbed externally. In QFT (and in STS below), where the emphasis is on dynamics, the preferred GF is instead the ground-state-to-ground-state matrix element. It is obtained from the partition function in the limit of infinitely long evolution (and, if necessary, Wick rotation), where only the eigenstate with the lowest eigenvalue contributes. This eigenstate is then declared the (global) ground state, and the GF takes the form of the corresponding ground-state-to-ground-state matrix element.

There are two additional reasons why the ground-state-to-ground-state matrix element is preferable in STS as a GF. First, when the ground-state eigenvalue is complex~\footnote{This situation can be interpreted as a spontaneous breakdown of pseudo-time-reversal symmetry.}, the partition function exhibits not only exponential growth as in Eq.~(\ref{Eq:Growth_pressure}), but also oscillations. In this case, defining a GF in terms of a quantity that crosses zero intermittently would be problematic. Second, unlike traces over all uncorrelated eigenstates, the state-to-state matrix element has a clear interpretation as an instance of stochastic evolution that starts in the infinite past and ends in the infinite future.

Because of TS, there are at least two eigenstates \footnote{There are four such eigenstates in case when the pseudo-time-reversal symmetry is broken and the ground state eigenvalue is complex.} with the real part of their eigenvalues equal $-\Delta$ with $\Delta$ from Eq.(\ref{Eq:Growth_pressure}). We can now pick any of these eigenstates as a ground state, $|g\rangle$. This state can be looked upon as a representative of self-sustained dynamics -- the infinitely long unperturbed dynamics of the DS. 

One can now introduce the external probing fields, $\zeta_a, a=1,...,n$, coupled to the model via a set of operators $O^a(\Phi)$. The GF is then expressed as,
\begin{subequations}
\label{Eq:GF}
\begin{eqnarray}
&e^{G(\zeta)} = \langle g| \mathcal{T} e^{\int d\tau \zeta_a(\tau)O^a(\hat\Phi(\tau))} |g\rangle = 
\end{eqnarray}
with $O^a(\hat\Phi(\tau))$ being the corresponding operators in the Heisenberg representation (\ref{Eq:HeisenbergRepresentation}). The same expression for GF in the partintegral representation is  
\begin{eqnarray}
= e^{TH_g} \iint \bar\Psi_g(T)  e^{\int d\tau\zeta_aO^a(\Phi)+S(\Phi)} \Psi_g(0)D\Phi,
\end{eqnarray}
\end{subequations}
where $\bar\Psi_g$ and $\Psi_g$ are the bra and ket of the ground state. The factor $e^{TH_g}$ is added to compensate the factor $e^{-TH_g}$ coming from the evolution of the unperturbed ground state. Due to this additional factor, $G(0) =1$. The exponential factor can be omitted if one adopts the convention that all eigenvalues are defined relative to the ground-state eigenvalue.

The description of dynamics provided by this GF is not restricted in any way by the fact that we chose one eigenstate over other fasterest growing eigenstates as a ground state, because during the evolution the system can visit any other state and explore the entire Hilbert space.

\color{darkBlue}
\section{Effective Field Theory}
\label{Sec:EFT}

\subsection{Higher-dimensional models}
\label{Sec:HigherDimensionalModels}

In this section, we discuss some aspects of the effective field theory (EFT) under the condition of spontaneously broken TS in higher-dimensional models. Accordingly, we shift our focus to stochastic partial differential equations (SPDEs). To avoid complications unrelated to the goals of this paper, we consider a model whose spatial part of the base space is a flat $d-1$-dimensional manifold, say, a torus. The base-space coordinate becomes $t\to (t,r)\equiv x$, and the fundamental variable of the model $\varphi(t)\to\varphi(x)$. 

The model is still defined by the action in Eqs.(\ref{Eq:Action}), with the only adjustments in Eqs.(\ref{Eq:SUSY}) and (\ref{Eq:QBar}) being $dt\to d^{d}x$, together with the remark that the vector fields entering the definition of the SDE in Eq.~(\ref{Eq:SDE_only}) contain spatial derivatives, as is the case, for example, in hydrodynamical equations.

In turning to higher-dimensional theories, we first point out that everything said above in the (1+0)-dimensional setting applies directly to lattice versions and/or numerical implementations of SPDEs, in which the spatial part of the base space is discrete, so that the dimension of the phase space remains finite, albeit very large. From this perspective, SPDEs may be regarded as infinite-dimensional extensions of the 1+0 models discussed above, with most of the conclusions carrying over unchanged. At the same time, the richer structure of the base space in SPDEs introduces qualitatively new features such as solitons.

It must also be pointed out, however, that passing to the strict continuum limit of the lattice may introduce additional subtleties of primarily theoretical importance. We sweep those under the carpet, in accordance with the qualitative and in many ways speculative character of the discussion in this section, whose primary goal is to draw attention to an important and challenging problem: uncovering the hidden spatiotemporal sctructure in chaos, whose existence is suggested by interpreting chaos as order associated with spontaneously broken TS and whose nature may ultimately turn out to be topological. Establishing such an "order in chaos" on a rigorous footing would constitute a major advance with potentially far-reaching implications across many scientific disciplines.

\subsection{Goldstone theorem and 1/f noise}
\label{Sec:GoldstoneTheorem}

The energy momentum tensor of the model is $\mathcal Q$-exact,
\begin{eqnarray}
    T_{\alpha\beta}(x) = \{\mathcal{Q},t_{\alpha\beta}(x)\},\label{Eq:Q_exact_EMT}
\end{eqnarray}
with the time-time component being the spatial density of the SEO discussed in the 1+0 dimensional setting, $\hat H = \int d^{d-1}r T_{00}(r) = \{\mathcal{Q},\int d^{d-1}rt_{00}(r)\}$, and $t_{00}$ being the spatial density of the quantity $\bar d$ in Eq.(\ref{Eq:QBar}). 

Because the TS is spontaneously broken in EFT (c.f., Eq.(\ref{Eq:TS_breaking_SDE})),
\begin{eqnarray}
\langle \{\mathcal{Q},t_{00}(x)\}\rangle = h \ne 0,\;  \text{Re} h <0,\label{Eq:TS_breaking}
\end{eqnarray}
where $h$ is the spatial density of the ground state eigenvalue and the brackets denote averaging in the sense defined in Eq.(\ref{Eq:GF}).

One may now invoke the standard Ward-identity argument demonstrating the existence of long-range excitations in the model. Starting from the path-integral representation of,
\begin{eqnarray}
\langle t_{00}(x') \rangle = 0,\label{Eq:WardIdentity_0}
\end{eqnarray}
one performs a position-dependent transformation $\delta \Phi(x)= \epsilon(x)\{\mathcal{Q},\Phi(x)\}$ under which $\delta t_{00}(x') = \epsilon(x') \{\mathcal{Q}, t_{00}(x') \}$ and $\delta S = \int d^d x (\partial_\mu \epsilon(x)) j^Q_\mu(x) $, where $j^Q_\mu$ is the conserved current associated with TS. One then chooses a spacetime region $V$ containing the point $x'$ and takes $\epsilon(x)$ to be a small constant inside $V$ and zero outside. This yields the integral form of the Ward identity, 
\begin{eqnarray}
   \oint_{\partial V} dS_\mu \langle j_\mu^{Q}(x) t_{00}(x')\rangle = h,\label{Eq:WardIdentity}
\end{eqnarray}
where $d S$ is a surface element of the boundary of $V$. 

The left-hand side remains equal to the constant $h$ regardless of how far the boundary is taken from $x'$. This indicates the existence of long-range excitations capable of propagating over arbitrarily large distances, implying that they are gapless. Accordingly, they must fall off algebraically rather than exponentially. The precise nature of these excitations is model specific. 

Eq.~(\ref{Eq:WardIdentity}) is directly related only to the fermionic excitations, reflecting the fermionic nature of TS. At the same time, it is understood that each fermionic branch must be accompanied by a bosonic counterpart since all non-supersymmetric states remain doubly degenerate even when TS is spontaneously broken. The existence of such a bosonic branch reflects the fact that TS is a part of the larger $\mathfrak{gl}(1|1)$ symmetry algebra.

Unless a physical observable/operator is carefully designed, it will be coupled to this gapless bosonic branch. Indeed, it may not be easy to come up with a perturbation of the lattice structure of a solid that is not coupled to transverse sound -- the Goldstone mode of crystallization. Therefore, the correlator of a generic physical observable will exhibit power-law behavior in the long-wavelength limit. This provides an explanation for 1/f noise -- the power-law temporal and spatial correlations characteristic of chaotic dynamics that have long resisted theoretical explanation.

In other words, the EFT must become scale invariant in the long-wavelength limit or, setting aside once again the associated theoretical subtleties, a conformal field theory (CFT) - the result foretold by Polyakov in the context of hydrodynamical chaos \cite{Polyakov_1993}. One refinement of this picture from STS is that the resulting EFT may in fact be logarithmic conformal rather than ordinary conformal, owing to the non-semisimplicity of the representation category of $\mathfrak{gl}(1|1)$ \cite{Schomerus_Saleur_2006}.

It should also be emphasized that this Goldstone scenario of 1/f noise cannot, by itself, account for the butterfly effect (BE) -- the effectively infinite memory of initial conditions that is characteristic of chaos. Unlike the power-law correlations associated with 1/f noise, which ultimately fall off with time, the memory underlying the BE persists indefinitely. This distinction points to the possibility that the BE may be rooted in a topological mechanism. At the end of this section, we will presents several speculative arguments supporting this conjecture.

\subsection{$GL(1|1)$ Ginzburg-Landau Theory}\label{Sec:GLtheory}

EFT of the Ginzburg-Landau (GL) type (see, \emph{e.g}, Ref. \cite{RFT_SSB_book}) is a powerful concept that must be applicable to many models exhibiting chaotic dynamics. One general type of chaotic dynamics, however, is particularly well suited for this approach. Namely, as discussed in Appendix \ref{Sec:Example_Model}, there are two major types of chaotic dynamics in STS with one being the noise-induced or instantonic chaos. It represents a broad, multidisciplinary dynamical phenomenon commonly referred to in the Literature as the edge-of-chaos~\cite{Lewin_1999,Crutchfield_2012,Crutchfield_2003} or self-organized criticality~\cite{Bak_1987,Watkins_2016,Aschwanden_2011_book,Meisel_2012_seizures}. Self-sustained dynamics in this phase is dominated by noise-induced instantons such as earthquakes, solar flares, neuroavalanches etc. This dynamics can be described as a fluid of solitons that are being created/annihilated via antiinstantons/instantons. One of the simplest models exhibiting this type of dynamics is discussed in Appendix \ref{Sec:Example_Model}. 

In chaotic solitonic fuilds there is a scale on which we can speak of solitons as fundamental units of dynamics and introduce slow fields and/or order parameters representing them. At this scale, the model can be defined in terms of these fields only with the corresponding action being the GL functional, $S$,
\begin{eqnarray}
&e^{G} = \iint e^{ \tilde S(\bar\psi,\psi )} {\mathcal{D}}\bar\psi{\mathcal{D}}\psi.\label{Eq:EFT_deinition}
\end{eqnarray}
Here, $G$ is the generating functional introduced in Eq.(\ref{Eq:GF}) with zero probing fields, $\psi = \{\psi_i| i=1,...,k\}$, is a collection of slow fields and/or order parameters representing the condensed nonlinear entities such as solitons, $\bar\psi = \{\bar\psi_i| i=1,...,k\}$ is their "anti-matter" fields, the boundary conditions are assumed periodic and antiperiodic for bosons and fermions, respectively. The effective nature of this theory is manifested in the interpretation of $\tilde S$ as the result of integrating out the fast degrees of freedom in the spirit of the Wilsonian renormalization-group approach.

As discussed in Sec.\ref{Sec:GL(1_1)_EFT}, EFT must have a global internal $GL(1|1)$ symmetry. Therefore, each slow field must be a (typical) irreducible representation of $GL(1|1)$, all of which are two-dimensional,
\begin{eqnarray}&
    \psi_i = \left( \begin{array}{c}
         \phi_i  \\
         \eta_i
    \end{array} \right),
\end{eqnarray}
with $i$ being the index enumerating the fields.

The wavefunctions of the EFT are functionals of $\psi$'s, $|\Psi\rangle \equiv \Psi(\psi)$, and the TS from Eq.(\ref{Eq:TheAlgebra}) can be defined on this Hilbert space as,
\begin{subequations}\label{Eq:New_algebra}
\begin{eqnarray}
   & \hat Q = \int d^{d-1} r \sum\nolimits_{i} \hat q_i,\; \hat q_i = \eta_i \frac{\delta}{\delta\phi_i },
\end{eqnarray}
The unity operator can be given as,
\begin{eqnarray}
    &\hat E = \int d^{d-1} r \sum\nolimits_{i}\hat e_i,\; 
    \hat e_i = e_i(\phi_i\frac{\delta}{\delta\phi_i} + \eta_i\frac{\delta}{\delta\eta_i}),\label{Eq:New_Unity}
\end{eqnarray}
which is essentially the operator of the total number of all particles in the wavefunction and constant $e_i$ parametrizes typical representations of $GL(1|1)$. One way to interpret this parameter is that the corresponding field may represent a tandem of "elementary particles". Under the conditions of the spontaneous symmetry breaking, this total number of particles is marcoscopically large and its fluctuations can be neglected which turns $\hat E$ into the unity operator up to an constant factor.

Using Eqs.(\ref{Eq:New_Unity}) and (\ref{Eq:TheAlgebra}), one concludes that 
\begin{eqnarray}
    &\hat Q^T = \int d^{d-1} r\sum\nolimits_{i}\hat q^T_i,\; \hat q^T_i= e_i \phi_i\frac{\delta}{\delta \eta_i},\\
    &\hat K = \int d^{d-1} r \sum\nolimits_{i}\hat k_i,\; \hat k_i=  \phi_i\frac{\delta}{\delta \phi_i} - \eta_i\frac{\delta}{\delta \eta_i},
\end{eqnarray}
\end{subequations}
As compared to the original model, the slow fields of EFT are linear which is the reason why the explicit form of $\mathsf{\hat Q}^T$ becomes available.

To sum it up, the representations of the $\mathfrak{gl}(1|1)$ provided by the slow fields,
\begin{eqnarray}
    &\hat e_i= e_i \hat 1_{2\times2}, \hat q_i = \hat \sigma^+, \hat q^T_i = e_i\hat \sigma^-, \hat k_i = \hat\sigma_z/2,\label{Eq:representation_slow_fields}
\end{eqnarray}
where $\sigma$'s are Pauli matrices. For a $\hat g_i = e^{ a^\alpha \hat r_{i,\alpha}}\in GL(1|1)$, where $\hat r_i = (\hat e_i, \hat q_i, \hat q^T_i, \hat k_i)$, 
\begin{eqnarray}
    \psi_i = \hat g_i \psi_i.
\end{eqnarray}
In addition, one must also introduce the antimatter fields,
\begin{eqnarray}
    &\bar\psi_i = \left( \begin{array}{c}
         \bar\phi_i  \\
         \bar\eta_i
    \end{array} \right)^{sT}=  (\bar\phi_i , -\bar\eta_i),\label{Eq:antifields}
\end{eqnarray}
with $sT$ denoting supertranspose, which are dual representations transforming as,
\begin{eqnarray}
    \bar\psi_i = \bar\psi_i \hat g^{-1}_i,
\end{eqnarray}
so that $\bar\psi_i \psi_i$ is a $GL(1|1)$-invariant. This fact can be used to establish the corresponding transformation rule for the antimatter field as well as the path-integral version of Eqs.(\ref{Eq:New_algebra}) that combine the transformation rule for both matter and antimatter fields. For example, 
\begin{eqnarray}
   & \mathcal{Q} = \sum\nolimits_i \mathcal{Q}_i, \mathcal{Q}_i = \eta_i \frac\delta{\delta\phi_i} + \bar\phi_i \frac\delta{\delta\bar\eta_i}.
\end{eqnarray}
The GL functional, or rather its density, must contain only $GL(1|1)$-invariant terms such as 
\begin{eqnarray}
&\bar\psi_i\hat M_i\psi_i = \bar \phi_i \hat M_i\phi_i - \bar\eta_i \hat M_i\eta_i = \{\mathcal{Q}, \bar\eta_i \hat M_i \phi_i \},\label{Eq:GaussianPieces}
\end{eqnarray}
where $\hat M_i = M_i(\hat \partial)$ with $M_i$ being a function of partial spatio-temporal derivatives. 

The question of possible structures of higher-order $GL(1|1)$-invariant terms is related the Clebsch-Gordan coefficients of this supergroup, which is a mathematically heavy topic \cite{Stoilova_2010_Clebsch_Gordan} that goes beyond the scope of this paper. We just want to point out that the GL functional must also be $\mathcal Q$-exact just like the action of the original theory. Therefore, it must contain only $\mathcal Q$-exact terms, such as those in Eq.(\ref{Eq:GaussianPieces}) and their products,
\begin{eqnarray}
    &\nonumber &(\bar \psi_i \hat M_i\psi_i) \bar \psi_j \hat M_j \psi_j = \{\mathcal{Q}, (\bar\eta_i \hat M_i \phi_i )\bar\psi_j \hat M_j \psi_j \} ...
\end{eqnarray}

Summing this up, the GL functional density must be $GL(1|1)$-symmetric and $\mathcal Q$-exact, $\tilde S = \int d^dx \tilde L$,
\begin{eqnarray}
    &\tilde L = \sum\nolimits_i \bar\psi_i \hat M_{i} \psi_i + ... = \{ \mathcal{Q}, \sum\nolimits_i \bar\eta_i \hat M_{i} \phi_i + ...\},
\end{eqnarray}
where dots denote high-order terms which are assumed local in the spirit of Ginzburg-Landau approach. 

\subsection{Goldstone modes}
The simplest GL-EFT of this type has only one field, $\psi$, and a Mexican hat potential,
\begin{eqnarray}
   &\tilde L = \bar \psi \partial^2  \psi + \lambda (\bar\psi\psi - v^2)^2 -\lambda v^4\nonumber\\
   &=\{\mathcal{Q}, \bar \eta \partial^2 \phi + \lambda \bar \eta \phi (\bar\psi\psi - 2 v^2) \},\label{Eq:GL_theory}
\end{eqnarray}
where $\partial^2 = \partial_t^2-\partial_r^2$, the characteristic speed is assumed unity, and $\lambda$ and $v$ are some positive constants. 

Eq.(\ref{Eq:GL_theory}) is a good candidate for EFT of the example model discussed in Appendix \ref{Sec:Example_Model} (see Fig.\ref{Fig_1}). In this case, the order parameter fields, $\bar\psi$ and $\psi$, represent the condensed antiinstantons and instantons which are the processes of creation and annihilation of pairs of kink and antikinks. At interesting point here is that due to fixed velocity of solitons, the EFT is Lorenzian even though the original model is Galilean. This is probably a feature of a wide class of models in the noise-induced chaotic phase where solitons have a fixed velocity.

The minimum of Eq.(\ref{Eq:GL_theory}) is achieved at $\bar\psi\psi = v^2$, which corresponds to the spontaneous breakdown of TS because this operator is $\mathcal Q$-exact,
\begin{eqnarray}
   \langle \bar\psi\psi(x) \rangle =\langle \{\mathcal{Q}, \bar\eta\phi(x)\} \rangle = v \ne 0. \label{Eq:EFT_Q_breaking}
\end{eqnarray}

It also corresponds to the spontaneous breakdown of $GL(1|1)$. Indeed, a field configuration that minimizes the action, say, $\psi_0=(v,0)$ and $\bar \psi_0=(v,0)$, is invariant with respect to the following subgroup,
\begin{eqnarray}
    \hat g \psi_0=\psi_0, \bar\psi_0 \hat g^{-1} = \bar \psi_0, \;\hat g(\lambda)=\text{diag }(1,e^\lambda).
\end{eqnarray}
suggesting that $GL(1|1)$ is broken down to $GL(1)$. Accordingly, in neglect of fluctuations in $\bar\psi\psi$, the EFT is the nonlinear sigma model on $GL(1|1)/GL(1)$ (see, e.g., Ref. \cite{Sethi_1994}). 

The magnitude of $\bar\psi\psi$, whose gapped fluctuations around $v^2$ can be characterized as a "hard" mode, is only one bosonic field out of two bosonic and two fermionic fields that $\bar\psi$ and $\psi$ comprise. The remaining one bosonic and two fermionic variables can be viewed as coordinates on the vacuum manifold and/or target space. They can be linked to the remaining transformation of $GL(1|1)$ in the following manner,
\begin{eqnarray}
    \psi = \hat g \psi_0 = v \left(\begin{array}{c}
        e^\pi \\ \eta 
    \end{array}\right),
    \hat g =  \left( 
    \begin{array}{cc}
    e^\pi & \bar \eta \\
    \eta & 1 
    \end{array}\right),
\end{eqnarray}
and
\begin{eqnarray}
    \bar\psi = \bar\psi \hat g^{-1} = v \left(\begin{array}{c}
        e^{-\pi} (1 + e^{-\pi} \bar\eta \eta) \\ \bar \eta 
    \end{array}\right)^{sT}
\end{eqnarray}
where,
\begin{eqnarray}
    \hat g^{-1} =  \left( 
    \begin{array}{cc}
    e^{-\pi}(1 + e^{-\pi} \bar\eta \eta)& - e^{-\pi}\bar \eta  \\
    -e^{-\pi} \eta  & 1 + e^{-\pi} \eta\bar\eta 
    \end{array}\right),
\end{eqnarray}
so that the Gaussian part of the action becomes,
\begin{eqnarray}
   \tilde L \propto (\partial \pi)^2 - (\partial \bar \eta)(\partial \eta) + ... \label{Eq:Goldstone}
\end{eqnarray}
The fields $\pi, \bar\eta,\eta$ are the gapless Goldstone particles with linear dispersion, which reflects that the EFT is Lorenzian. 

A systematic way to account for fluctuations beyond the mean-field analysis may be a complex task. Here we only qualitatively touch upon this subject and in doing so one must first point out that STS has no temperature in the sense of statistical physics. The intensity of the noise is closer in its spirit to the kinetic energy as can be seen from Eq.(\ref{Eq:SEO_LieDerivatives}). The model can be forced to visit states other than the ground state only externally. In other words, there are only "quantum" or intrinsic fluctuations in STS and the thermodynamic temperature is always zero.

As in other nonlinear sigma models, the one-loop beta function is given by the Ricci tensor of the target space, which vanishes in our case. This suggests that the Goldstone modes must survive the fluctuations, or, rather, they will yield corresponding conformal blocks rendering the theory scale-invariant in the long-wavelength limit. 

The model also has a $\mathbb{Z}_2$ variable, $\varepsilon$, whose dynamics is associated with transitions between two disconnected components of $GL(1|1)/GL(1)$: $\bar\phi,\phi \leftrightarrow -\bar\phi,-\phi$. 

This introduces Ising-type dynamics into the picture, with two possible scenarios. When Ising-type ordering is possible, $\langle \varepsilon(x) \varepsilon(0) \rangle \stackrel{|x|\to\infty}{\to} const\ne 0$, the model can be viewed in the long-wavelength limit as living on a single branch of the target space so that the Goldstone modes should survive. In 1+1 dimensions, however, only a quasi-long-range-order is possible with power-law correlations, $\langle \varepsilon(x) \varepsilon(0) \rangle \stackrel{|x|\to\infty}{\to} 1/|x|^\alpha$. In this case, $\varepsilon$ itself becomes, or contributes to, the Goldstone excitation responsible for the emergent scale invariance of the long-wavelength theory.

In both scenarios, however, the model will exhibit scale-invariant behavior in the long-wavelength limit. This is in accordance with the general argument based on the Ward identity at the end of Sec.\ref{Sec:HigherDimensionalModels}.

\subsection{Supergauge-invariance of the generating functional}

The external supergauge field can be coupled to the model by switching to covariant derivatives,
\begin{eqnarray}
    &\hat M_i \to M_i(\hat D_i), \hat D_{i,\mu} = \partial_\mu + \hat{A}_{i,\mu}\; \hat{A}_{i,\mu} =\hat{A}^\alpha_\mu\hat r_{i,\alpha},
\end{eqnarray}
with $\hat r_i$'s from Eq.(\ref{Eq:representation_slow_fields}). Accordingly, $\tilde S \to \tilde S(A) = \int d^dx \tilde L(A),$ and the generating functional also becomes the functional of $A$,
\begin{eqnarray}
    &e^{G(A)} = \iint e^{\tilde S(A,\bar\psi,\psi)}D\psi D\psi.\label{Eq:G_A}
\end{eqnarray}
In the presence of nonzero background gauge field, the EFT is invariant under the local $GL(1|1)$ gauge transformations,
\begin{eqnarray}
\tilde S(A,\bar\psi g^{-1},g\psi) = \tilde S(A+g^{-1}\partial g,\bar\psi,\psi),\label{Eq:Invariance_at_A}
\end{eqnarray}
where $g$ is a position dependent $GL(1|1)$ transformation of the fields. Accordingly, the generating functional is $GL(1|1)$ gauge-invariant because any gauge transformation of $A$ can be compensated by the opposite transformation of the fields, provided that there are no anomalies which go beyond the scope of this paper. This argument can be symbolically expressed as,
\begin{eqnarray}
    &e^{G(A-g^{-1}\partial g)} = \iint e^{\tilde S(A-g^{-1}\partial g,\bar\psi,\psi)}D\bar\psi D\psi\nonumber \\
    &=\iint e^{\tilde S(A,\bar\psi g^{-1},g\psi)}D\bar\psi g^{-1} Dg\psi\nonumber \\
    &=\iint e^{\tilde S(A,\bar\psi,\psi)}D\bar\psi D\psi = e^{G(A)},\label{Eq:gauge_invariance}
\end{eqnarray}
where  Eq.(\ref{Eq:Invariance_at_A}) and $GL(1|1)$-invariance of the integration measure, $D\bar\psi D\psi = D\bar\psi g^{-1} Dg\psi$, are utilized. 

The difference between the broken and unbroken symmetry cases is not whether $G(A)$ is gauge invariant. It remains gauge invariant even when the symmetry is spontaneously broken and the external gauge field is Higgsed, as explained in Appendix \ref{Sec:AppendixB}. The difference is that $G(A)$ becomes highly nonlocal in the symmetry-broken phase.

\subsection{EFT as TFTs}
\label{Sec:EFT_TFT}

With the aid of the concept of cohomological reduction, it is argued in Refs.~\cite{Candu_2010,Quella_2013} that symmetry-broken phases of supersymmetric models must contain nontrivial topological sectors and may even be described by TFTs. This observation is directly relevant to the problem of the EFT of STS and the possible topological character of the butterfly effect. Within the framework adopted in this paper, this line of reasoning can be complemented in the following manner.

We first notice that the action of the EFT is $Q$-exact even in the presence of external field provided that the definition of TS is extended to include the transformations of $A$,
\begin{eqnarray}
    &\tilde L(A) = \{ \mathcal{Q}', \sum\nolimits_i \bar\eta_i M_{i}(\hat D_i) \phi_i + ...\},
\end{eqnarray}
where,
\begin{eqnarray}
    \mathcal{Q}' = \mathcal{Q} + \mathcal{Q}_A,\label{TS_TOTAL}
\end{eqnarray}
with
\begin{eqnarray}
\mathcal{Q}_{A} = \int d^{d}x\sum\nolimits_{\mu} ( A_\mu^{k}\delta/\delta A_\mu^{q} + A_\mu^{q^T}\delta/\delta A_\mu^{e}).\label{TS_A}
\end{eqnarray}
This implies that the GF in Eq.(\ref{Eq:G_A}) is obtained by partial integration of fields in a model with a $Q$-exact action. The $Q$-exactness is a robust property that is difficult to destroy, which, in turn, suggests that
\begin{eqnarray}
    G(A) = \{ \mathcal{Q}_{A} , \omega(A) \},\label{Eq:Q_exact_action}
\end{eqnarray}
with $\omega$ being unspecified functional of $A$. 

If Eq.(\ref{Eq:Q_exact_action}) could be proved, there would appear a possibility to construct a cohomological TFT which is dual -- in a sense  -- to the EFT, which, in turn, would demonstrate the topological character of the BE. More specifically, some of the following matrix elements can turn out to be topological invariants,  
\begin{eqnarray}
\iint \mathcal{O}_{i_1} ... \mathcal{O}_{i_k} e^{\{\mathcal{Q}_A, \omega(A)\}} D A,
\label{Eq:TopInv}
\end{eqnarray}
where $O_i$ are some $Q$-closed operators, $[\mathcal{Q}_{A},\mathcal{O}_i] = 0$, which are not $Q$-exact, ${\mathcal O}_{i} \ne [{\mathcal Q}_{A}, o], $ $\forall o$. One type of such operators is Wilson loops (see Fig.\ref{Fig_1}b),
\begin{eqnarray}
    \mathcal{W}_{C} = \text{sTr } P e^{\oint_C \hat A_\mu(x) dx^\mu},\label{Eq:WilsonLoops}
\end{eqnarray}
where $P$ denotes path ordering along a closed contour $C$.

The computation of Wilson loops involving superconnections is a known problem in high-energy physics (see, e.g., Refs. \cite{Maldacena_1998,Lee_2010} and references therein), with the result most relevant to our case being the relation between Wilson loops of $GL(1|1)$ and the Alexander polynomial \cite{Rozansky_1992}. Hopefully, some of these results can be adapted to the EFT of the TS-broken phase of STS.

\subsubsection{Holography and EFTs}

Eq.(\ref{Eq:Q_exact_action}) can be further supported with the help of the Anti-de Sitter (AdS)/CFT correspondence~\cite{Maldacena_1999,Gubser_1998,Witten_1998,Natsuume_2016_book,Penedones_2016,Kaplan_2016_book}. There, nonlocal response of the model originates from the propagation in additional dimension having the meaning of a scale variable.

More specifically, the presence of power-law correlations associated with Goldstone particles, suggested by the Ward identity in Eq.(\ref{Eq:WardIdentity}), is the reason why AdS/CFT correspondence is applicable to spontaneous symmetry breaking \cite{Kaplan_2016_book} such as superconductiors \cite{Ren_2010,Alkac_2017}. Moreover, chaotic solitonic fluids, which N-phase dynamics essentially is, fall into the class of soliton fluids to which AdS/CFT is also applicable~\cite{Ker_nen_2009,Ker_nen_2010}. Therefore, we find it reasonable to believe that AdS/CFT correspondence must provide a reasonable approximation in our case too.

The GKP-Witten relation provides the following approximation to the GF,
\begin{eqnarray}
    G(A) =  \min_{\mathcal A, \;\mathcal{A}|_{B}=A}\mathcal{S}(\mathcal{A}),\label{Eq:AdS}
\end{eqnarray}
where the minimization in the r.h.s. is over the bulk gauge field $\mathcal A$ subject to the condition at the conformal boundary of AdS, $B$, that $\mathcal A$ equals (perhaps up to a function of distance to the boundary) to $A$. Once again, the temperature in STS is always zero so that the backgrounds with a blackhole should not be expected.

The bulk action may contain the Yang-Mills term which is the key ingredient of the holographic dual of soliton fluids \cite{Ker_nen_2009,Ker_nen_2010}. The action may also contain the Chern-Simons term which is a natural addition for 2D systems in which it would represent left/right movers. This term can also be expected in models with broken pseudo-time-reversal symmetry (see Sec.\ref{Sec:etaTbreaking}) when the EFT must have two subparts corresponding to two vacua with complex conjugate "energies". The action can also contain the Higgs term which may appear when the gauge field is coupled to a condensate.

It can be shown that the Yang-Mills, Chern-Simons, and the Higgs terms are all $\mathcal{Q}_A$-exact, \emph{e.g.}, the Lagrangian density of the Higgs term,
\begin{eqnarray}
    \mathcal{L}_\text{H}(\mathcal{A}) = \text{sTr} \hat{\mathcal{A}}\hat{\mathcal{A}} = \{\mathcal{Q}_{A}, \mathcal{A}^{q}\mathcal{A}^{e} \}. 
\end{eqnarray}
with $\text{sTr} \hat{\mathcal{A}}\hat{\mathcal{A}}  = \mathcal{A}^{k}\mathcal{A}^{e} - \mathcal{A}^{q}\mathcal{A}^{q^T}$ being a supertrace. This observation suggests the existence of a broad class of models whose AdS/CFT duals can be used to realize the scenario outlined in Sec.\ref{Sec:EFT_TFT}.

To conclude the discussion in this section let us say a few words about the AdS/CFT dual of the example model in Appendix \ref{Sec:Example_Model} (see Fig.\ref{Fig_1}c). Its EFT can be given by the Lagrangian density 
\begin{eqnarray}
    &\tilde L = \sum_{a=\pm} \bar\psi_a \partial_{\pm} \psi_a + ...
\end{eqnarray}
where the fields represent the left/right moving kinks and antikinks and dots represent instantonic processes of creation and annihilation of kink-antikink pairs. The corresponding coupling must be sufficiently strong to dynamically forbid kinks and antikinks to travel through each other. The following quasi-long-range order parameters must appear, $\langle \psi _+ \psi_-\rangle, \langle \bar\psi _+ \bar \psi_-\rangle \ne 0$.

Using $AdS_3/CFT_2$ dictionary, the bulk action must have two $GL(1|1)$ gauge fields, $\mathcal{A}_\pm$, representing the $GL(1|1)$ currents of the corresponding left and right movers, and an instanton quasi-condensate wavefunction, $\Phi$, which is the operator of creation and annihilation of kink-antikink pairs in the boundary theory,  $\Phi\sim\langle \psi _+ \psi_-\rangle, \bar\Phi\sim\langle \bar\psi _+ \bar \psi_-\rangle$. Symbolically, the dual can be expressed as \footnote{This model can also be viewed as an AdS/CFT dual of the $GL(1|1)$ Wess–Zumino–Witten model \cite{Schomerus_Saleur_2006} suggesting that our example model must be related to it.},
\begin{eqnarray}
    \mathcal{L}(\mathcal{A}) \propto \mathcal{L}_{SC}(\mathcal{A}_+) - \mathcal{L}_{SC}(\mathcal{A}_-) + |D \Phi|^2 + V(\bar \Phi\Phi).\label{Eq:WZW_dual}
\end{eqnarray}
Here, the covariant derivative $D\Phi = (\partial \Phi + \mathcal{A}_+\Phi + \Phi \mathcal{A}_-^{sT}) $ couples the condensate to the left and right movers on equal footing, and $V(\bar \Phi\Phi)$ must be a simple quadratic potential because in $1+1$ dimensions (and zero temperatures) only a quasi-condensate exists corresponding on the CFT side to a primary field with the conformal weight related to the bulk mass of $\Phi$. 

The symmetric combination of the gauge fields, $\mathcal{A}_+ + \mathcal{A}_-$, must be (quasi-)Higgsed due to interaction with $\Phi$, while the antisymmetric combination, $\mathcal{A}_+ - \mathcal{A}_-$, must remain long-ranged and topological, so that one can expect that the corresponding Wilson loops (see Fig.\ref{Fig_1}c) are related to the Alexander polynomial \cite{Rozansky_1992}.

We conclude this section by emphasizing that, beyond the technical aspects associated with EFTs of STS, it is equally important to understand how objects such as topological invariants should be interpreted within the conventional framework of chaotic dynamics.

\color{black}
\section{Conclusion}
\label{Sec:Conclusion}

In this paper, we have addressed the apparent contradiction that, on the one hand, TFTs have no local excitations, while, on the other hand, within the Parisi–Sourlas–Baulieu–Grossman framework, TFTs describe SDEs where local excitations are intrinsic. The origin of this inconsistency lies in a common misinterpretation of the Parisi–Sourlas functional as the partition function of the SDE. In fact, it is a topological invariant -- the Witten index -- which serves as a representative of the partition function of the noise. The latter is a static probabilistic entity with no intrinsic dynamics.

The partition function of the DS itself is obtained by imposing anti-periodic boundary conditions on the ghosts, thereby rendering them physical. This step is justified provided one assigns a meaningful interpretation to the ghosts, for example as differentials used to track Lyapunov exponents in numerical experiments.

We have also emphasized that, from a more general perspective, an SDE does not define a partition function, a Witten index, or any other trace-like object. Rather, it generates noise-dependent trajectories whose action on the exterior algebra of phase space is encoded in the stochastic evolution operator. Unlike trace-like quantities, state-to-state matrix elements of the stochastic evolution operator admit a clear interpretation as instances of stochastic evolution. These include transient processes, or instantons, associated with transitions between (perturbative) supersymmetric ground states, as well as sustained dynamics corresponding to evolution from the (global) ground state to itself. The latter provides the natural starting point for constructing an effective field theory -- an important tool for uncovering long-range (or even topological) spatio-temporal order which is likely concealed in chaotic dynamics. \color{myblue} We discussed Ginzburg-Landau approach to the effective field theory and speculated on the potential usefulness of the concept of cohomological reduction and AdS/CFT correspondence in uncovering this "order in chaos".\color{black}

In view of the foundational role of the stochastic evolution operator, we have reviewed its relation to a few other objects encountered in stochastic dynamics.

\section*{Acknowledgments}
The author acknowledges initial support from the DARPA BAA “Physical Intelligence” program, and is especially grateful to Kang L. Wang for enabling this work. The author also thanks Massimiliano Di Ventra, Stam Nicolis, Savdeep Sethi, Gabriel Weiderpass, and Torsten Ensslin for discussions. Additional appreciation is extended to Robert Schwartz, Ben Israelii, Daniel Toker, Skirmantas Janusonis, Mohit Gupta, and Eugene Ingerman, all of whom positively influenced this work.
\appendix
\color{darkBlue}
\section{Example model}
\label{Sec:Example_Model}
At weak noises, there are two major types of chaos in STS corresponding to two different levels at which the TS is broken (see, e.g.,  Ref.\cite{Ovchinnikov_2016})~\footnote{In case of spontaneously broken bosonic symmetries, there is also a possibility of anomaly, which, however, is unlike to break a supersymmetry.}. The first is conventional chaos, in which TS is broken due to the non-integrability of the flow vector field in the SDE. In field theoretic language, this corresponds to the gaussian level of TS breaking. To see this, one recalls that for integrable flows, there is a global unstable manifold and the corresponding global supersymmetric ground state is its Poincare dual widened by a weak noise in transverse directions into a narrow Gaussian distribution. In chaotic or non-integrable flows, the global unstable manifold is not a topological manifold but a fractal structure called strange attractor and the above picture of the global ground state breaks down at the Gaussian level.

The second type is instantonic, or noise-induced, chaos, where the flow in integrable but the TS is spontaneously broken by noise-induced instantons or, more accurately, by the configurations of the noise-induced antiinstantons and instantons matching them. The instantnoic chaos may be qualitatively understood as the boundary of the conventional chaos broadened into a finite-width phase by the presence of noise. In the absence of noise, it collapses onto the critical boundary of the conventional chaos (see Fig.\ref{Fig_1}b). 

One of the simplest models that can be used to explore the noise-induced chaotic dynamics is the overdamped stochastic sine-Gordon equation \cite{sine_Gordon_Kink_Bath,Classics,Josephson_general,PhysRevB.54.1234}, or Frenkel-Kontorova equation \cite{Review_Frenkel_Kontorova}, with a non-potential driving discussed in Ref.[\onlinecite{Ovchinnikov_2024}]. It can also be viewed as a toy model in neurodynamics where it represents a ring of type-I neurons~\cite{Ovchinnikov_2020}. 

The spatial part of its basespace is a circle of circumference $R$ and the SDE is given by,
\begin{subequations}\label{sineGordon}\begin{eqnarray}
\partial_t \varphi(x) &=& \alpha - V'_{sG}(\varphi)(x) + \sqrt{2\Theta}\xi(x),
\label{SGE}
\end{eqnarray}\end{subequations}
where $\varphi\in \mathbb{S}^1$, $V_{sG} = \int dr \left( (\partial_r\varphi)^2/2 - \cos(\varphi)\right)$ is the sine-Gordon potential, $V'_{sG} = -\partial_r^2\varphi + \sin\varphi$, $\alpha$ is the non-potential driving, and $\overline{\xi(x)\xi(x')} = \delta^2(x-x')$.

At $\alpha < 0$, there is the stable vacuum solution $\varphi_0 = \sin^{-1}\alpha$. At $\alpha>1$, this solution disappears and dynamics is chaotic even in the deterministic limit. The noise-induced chaotic dynamics occurs at $1-\Delta(\Theta)<\alpha<1$, with $\Theta, \Delta(\Theta)<<1$. It consists of left-moving kinks and right-moving antikinks that are being created and annihilated in pairs in instantonic and antiinstantnoic processes (see Fig.\ref{Fig_1}c). 

The instantnoic processes of annihilation of two solitons is a deterministic process which does not require assistance from the noise. In contrast, the anti-instantonic process of creation of a kink–antikink pair occurs only due to the noise. Therefore, antinstantnoic processes along with the N-phase itself disappears in the deterministic limit, $\Theta\to0$.

To verify that TS is indeed broken in this example, one may examine the model’s partition function. It can be approximated using the observation that each multi-instanton configuration is uniquely specified by the spatiotemporal positions of its (anti)instantons. Neglecting perturbative corrections, which seems reasonable in the light of the boson-fermion cancellations in supersymmetric models, the partition function reduces to the integral over the moduli space of multi-instanton configurations,
\begin{eqnarray}
   & Z(T) \approx \sum_{i=0}^{\infty} \frac{1}{k!}(\int_{0}^T dt \int_0^R dr e^{-\Delta U/\Theta})^k \nonumber\\
   &  =\sum_{i=0}^{\infty} \frac{1}{k!}(-T H_g)^k = e^{-T H_g}, \; H_g = -R e^{-\Delta U/\Theta},
\end{eqnarray}
where $R$ is the length of the spatial circle, the factorial $k!$ accounts for the indistinguishability of (anti)instantons, and the Gibbs factor $e^{-\Delta U/\Theta}$ arises from antiinstanton processes corresponding to the creation of kink–antikink pairs, with $\Delta U$ denoting the "energy" needed to create a single kink-antikink pair, which vanishes at the border of the C-phase, $\Delta U(\alpha)\to0, \alpha\to1$. The resulting ground state eigenvalue $H_g$ has a negative real part and zero imaginary part, implying that TS is spontaneously broken while pseudo-time-reversal symmetry is unbroken.

An interesting observation about this model is that the solitons have a fixed velocity, so that the corresponding EFT must be Lorentzian even though the parental model is  Galilean. This is probably true for a whole class of models in which the noise-induced chaotic dynamics involves solitons with a fixed velocity.

In connection with the discussion in Sec. \ref{Sec:SUSY}, it is worth noting that the operator $\hat{\bar d}$ in Eq.(\ref{Eq:DExactH}) is nilpotent in this model, so that the second supersymmetry is known explicitly.

\section{Gauge-invariance and the Higgs Mechanism}
\label{Sec:AppendixB}
In the context of spontaneous symmetry breaking, it is often said that the corresponding gauge symmetry is broken by a non-gauge-invariant vacuum or condensate. If the condensate is treated as a dynamical field, however, one must integrate over all of its possible configurations, turning the theory into a nonlinear sigma model. This integration restores the gauge invariance. 

The simplest way to demonstrate this mechanism is to consider the $O(1)$ sigma model, say, of a superconductor, coupled to an external gauge field,
\begin{eqnarray}
    &G(A) = -\log \iint e^{ -\int (\partial \varphi - A)^2 d^d x } D\varphi \nonumber \\
    &= \int d^d x (A^2 - (\partial A) (\partial^2)^{-1}(\partial A))\nonumber\\
    &= \int \frac{d^d k}{(2\pi)^d} \left(A_\mu(k)(\delta_{\mu\nu}-k_\mu k_\nu/k^2))A_\nu(-k)\right),
\end{eqnarray}
where in the last line the space-time indices are explicitly given and the momentum space representation is used. The invariance with respect to gauge transformation $A_\mu(k)\to A_\mu(k) + k_\mu f(k)$ is manifest. 

The lesson from this simple example is that the dynamics of the condensate compensates for the apparent violation of gauge invariance introduced by the Higgs term through the emergence of an additional contribution that is highly nonlocal in space-time.
\color{black}
\bibliography{my_bibliography}

@article{Losev_2005,
author = {Losev, A.},
title = {Topological quantum mechanics for physicists.},
journal = {JETP. Lett.},
volume = {82},
pages = {335--342},
year = {2005},
}

@inproceedings{Biology1fNoise,
author = {Musha, T. and Mitsuaki, Y.},
title = {$1/f$ Fluctuations in Biological Systems},
booktitle = {Proceedings of the 19th Annual International Conference of the IEEE on Engineering in Medicine and Biology Society, Chicago, IL, USA, 30~October--2~November~1997},
pages = {2692--2697},
year = {1997},
}

@article{Mostafazadeh_2002,
author = {Mostafazadeh, A.},
title = {Pseudo-Hermiticity versus PT-symmetry II: A complete characterization of non-Hermitian Hamiltonians with a real spectrum.},
journal = {J. Math. Phys. },
volume = {43},
pages = {2814--2816},
year = {2002},
}

@article{Mostafazadeh_2002_1,
author = {Mostafazadeh, A.},
title = {Pseudo-Hermiticity versus PT-symmetry III: Equivalence of pseudo-Hermiticity and the presence of antilinear symmetries},
journal = {J. Math. Phys.},
volume = {43},
pages = {3944--3951},
year = {2002},
}

@article{Olenskoi_2004,
author = {Olemskoi, A.I. and Khomenko, A.V. and Olemskoi, D.A.},
title = {Field theory of self-organization},
journal = {Physica A},
volume = {332},
pages = {185--206},
year = {2004},
}

@article{Eckmann_Ruelle_1985_RMP,
author = {Eckmann, J.P. and Ruelle, D.},
title = {Ergodic theory of chaos and strange attractors.},
journal = {Rev. Mod. Phys.},
volume = {57},
pages = {617--656},
year = {1985}
}

@article{Ovchinnikov_2016,
author = {Ovchinnikov, I. V.},
title = {Introduction to supersymmetric theory of stochastics},
journal = {Entropy},
volume = {18},
pages = {108},
year = {2016}
}

@article{Bak_1987,
author = {Bak, P. and Tang, C. and Wiesenfeld, K.},
title = {Self-organized criticality: An explanation of the $1/f$ noise},
journal = {Phys. Rev. Lett.},
volume = {59},
pages = {381--384},
year = {1987}
}

@book{MirrorSymmetry,
author = {Hori, K. and Katz, S. and Klemm, A. and Pandharipande, R. and Thomas, R. and Vafa, C. and Vakil, R. and Zaslow, E.},
title = {Mirror Symmetry},
publisher = {American Mathematical Society and Clay Mathematics Institute},
address = {Cambridge, MA, USA},
year = {2003},
}

@article{Dijkgraaf_2010,
author = {Dijkgraaf, R. and Orlando, D. and Reffert, S.},
title = {Relating field theories via stochastic quantization},
journal = {Nucl. Phys. B},
volume = {824},
pages = {365--386},
year = {2010}
}

@book{BookHeartBrainNoise,
author = {Dana, S.K. and Roy, P.K. and Kurths, J. (Eds.)},
title = {Complex Dynamics in Physiological Systems: From Heart to Brain},
publisher = {Springer },
address = {Berlin/Heidelberg, Germany },
year = {2009},
}

@article{Gozzi_1993,
author = {Gozzi, E.},
title = {Stochastic and Non-Stochastic Supersymmetry},
journal = {Prog. Theor. Phys. Suppl.},
volume = {111},
pages = {115--150},
year = {1993},
}

@article{Gozzi_1990,
author = {Gozzi, E. and Reuter, M.},
title = {Classical mechanics as a topological field theory.},
journal = {Phys. Lett. B},
volume = {240},
pages = {137--144},
year = {1990},
}

@article{Gozzi_1994_Lyapunov,
author = {Gozzi, E. and Reuter, M.},
title = {Lyapunov exponents, path-integrals and forms},
journal = {Chaos Solitons Fractals},
volume = {4},
pages = {1117--1139},
year = {1994},
}

@book{Nakahara_book,
author = {Nakahara, M.},
title = {Geometry, Topology, and Physics},
publisher = {IOP Publishing},
address = {Bristol, UK},
year = {1990},
}

@article{Ruelle_2014,
author = {Ruelle, D.},
title = {Early chaos theory.},
journal = {Phys. Today},
volume = {67},
pages = {9--10},
year = {2014}
}

@article{Tailleur_2006,
author = {Tailleur, J. and T\"anase-Nicola, S. and Kurchan, J.},
title = {Kramers equation and supersymmetry},
journal = {J. Stat. Phys.},
volume = {122},
pages = {557--595},
year = {2006},
doi={https://doi.org/10.1007/s10955-005-8059-x}
}

@article{Parisi_Sourlas_1979,
author = {Parisi, G. and Sourlas, N.},
title = {Random magnetic fields, supersymmetry, and negative dimensions},
journal = {Phys. Rev. Lett.},
volume = {43},
pages = {744--745},
year = {1979}
}

@article{Witten_1988_1,
author = {Witten, E.},
title = {Topological sigma models},
journal = {Commun. Math. Phys.},
volume = {118},
pages = {411--449},
year = {1988}
}

@article{ZinnJustin_1986,
author = {Zinn-Justin, J.},
title = {Renormalization and stochastic quantization},
journal = {Nucl. Phys. B},
volume = {275},
pages = {135--159},
year = {1986}
}

@article{Frenkel_2007,
author = {Frenkel, E. and Losev, A. and Nekrasov, N.},
title = {Notes on instantons in topological field theory and beyond},
journal = {Nucl.~Phys.~B},
volume = {171},
pages = {215--230},
year = {2007},
doi={https://doi.org/10.1016/j.nuclphysbps.2007.06.013}
}

@article{CecottiGirardello1983,
author = {Cecotti, S. and Girardello, L.},
title = {Stochastic and parastochastic aspects of supersymmetric functional measures: A~new non-perturbative approach to supersymmetry.},
journal = {Ann. Phys.},
volume = {145},
pages = {81--99},
year = {1983},
doi={https://doi.org/10.1016/0003-4916(83)90172-0}
}

@article{CecottiGirardello1982,
  author  = {Cecotti, S. and Girardello, L.},
  title   = {Functional measure, topology and dynamical supersymmetry breaking},
  journal = {Physics Letters B},
  volume  = {110},
  number  = {1},
  pages   = {39--43},
  year    = {1982},
  doi     = {10.1016/0370-2693(82)90947-9}
}

@article{CecottiGirardello1984,
  author  = {Cecotti, Sergio and Girardello, Luciano},
  title   = {A supersymmetry anomaly and the fermionic string},
  journal = {Nuclear Physics B},
  volume  = {239},
  number  = {2},
  pages   = {573--582},
  year    = {1984},
  doi     = {10.1016/0550-3213(84)90263-3}
}

@article{Ruelle_2002,
author = {Ruelle, D.},
title = {Dynamical zeta functions and transfer operators},
journal = {Not. AMS},
volume = {49},
pages = {887--895},
year = {2002}
}

@book{Lewin_1999,
author = {Lewin, R.},
title = {Complexity: Living on the Edge of Chaos},
publisher = {University of Chicago Press},
address = {Chicago, IL, USA},
year = {1999},
}

@article{Deotto_2003,
author = {Deotto, E. and Gozzi, E. and Mauro, D.},
title = {Hilbert space structure in classical mechanics. I},
journal = {J. Math. Phys., },
volume = {44 },
pages = {5902--5936},
year = {2003},
doi={https://doi.org/10.1063/1.1623333}
}

@article{Kleinert_Shabanov_1997,
author = {Kleinert, H. and Shabanov, S.V.},
title = {Supersymmetry in stochastic processes with higher-order time derivatives},
journal = {Phys.~Lett.~A},
volume = {235},
pages = {105--112},
year = {1997},
doi={https://doi.org/10.1016/S0375-9601(97)00660-9}
}

@article{Niemi_1986,
author = {Niemi, A.J. and Pasanen, P.},
title = {Topological $\sigma$-model, Hamiltonian dynamics and loop space Lefschetz number},
journal = {Phys. Letts. B},
volume = {386},
pages = {123--130},
year = {1996},
doi={https://doi.org/10.1016/0370-2693(96)00941-0}
}

@article{Witten_1988,
author = {Witten, E.},
title = {Topological quantum field theory},
journal = {Commun. Math. Phys.},
volume = {117},
pages = {353--386},
year = {1988}
}

@article{Drummond_2012,
author = {Drummond, I.T. and Horgan, R.R.},
title = {Stochastic processes, slaves and supersymmetry.},
journal = {J. Phys. A},
volume = {45},
pages = {095005},
year = {2012}
}

@book{Aschwanden_2011_book,
author = {Aschwanden, M.},
title = {Self-Organized Criticallity in Astrophysics: Statistics of Nonlinear Processes in the Universe},
publisher = {Springer},
address = {Berlin/Heidelberg, Germany},
year = {2011},
}

@book{Kapitaniak_1990,
author = {Kapitaniak, T.},
title = {Chaos in Systems with Noise},
publisher = {World Scientific},
address = {Singapore, Singapore},
year = {1990},
}

@article{Parisi_Sourlas_1982,
author = {Parisi, G. and Sourlas, N.},
title = {Supersymmetric field theories and stochastic differential equations},
journal = {Nucl. Phys. B},
volume = {206},
pages = {321--332},
year = {1982}
}

@article{Labastida_1989,
author = {Labastida, J.M.F.},
title = {Morse theory interpretation of topological quantum field theories},
journal = {Commun. Math. Phys.},
volume = {123},
pages = {641--658},
year = {1989}
}

@article{Gilmore_1988,
author = {Gilmore, R.},
title = {Topological analysis of chaotic dynamical systems},
journal = {Rev. Mod. Phys.},
volume = {70},
pages = {1455--1529},
year = {1998},
}

@book{Kunita_1997_book_SDEs,
author = {Kunita, H.},
title = {Stochastic Flows and Stochastic Differential Equations},
publisher = {Cambridge University Press},
address = {Cambridge, UK},
year = {1997},
}

@article{Preis_2011,
author = {Preis, T. and Schneider, J.J. and Stanley, H.E.},
title = {Switching processes in financial markets},
journal = {Proc. Natl. Acad. Sci. USA},
volume = {108},
pages = {7674--7678},
year = {2011},
}

@article{Bouya_2013,
author = {I. Bouya and E. Dormy}, 
title = {Revisiting the ABC flow dynamo},
journal = {Physics of Fluids},
volume = {25},
pages = {037103},
year = {2013},
}

@article{Ovchinnikov_2018,
	year = 2018,
	month = {jun},
	publisher = {{IOP} Publishing},
	volume = {2},
	number = {6},
	pages = {065008},
	author = {Ovchinnikov, I. V. and Sun, Y. and En{\ss}lin, T. A.  and Wang, K.L. },
	title = {Supersymmetric theory of stochastic {ABC} model},
	journal = {Journal of Physics Communications}
}

@article{Baulieu_1988,
author = {Baulieu, L. and Grossman, B.},
title = {A topological interpretation of stochastic quantization},
journal = {Physics Letters B},
volume = {212},
number = {3},
year = {1988},
pages = {351}
}

@article{Baulieu_1989_1,
author = {Baulieu, L. and Singer, M.},
title = {The topological sigma model},
journal = {Comms. in Math. Phys.},
volume = {125},
number = {2},
year = {1989},
pages = {227}
}

@article{Ensslin_2016,
  title = {Kinematic dynamo, supersymmetry breaking, and chaos},
  author = {Ovchinnikov, I.V. and En\ss{}lin, T.A.},
  journal = {Phys. Rev. D},
  volume = {93},
  issue = {8},
  pages = {085023},
  numpages = {10},
  year = {2016},
  publisher = {American Physical Society}
}

@article{Uthamacumaran_2021,
title={A review of dynamical systems approaches for the detection of chaotic attractors in cancer networks},
author={Uthamacumaran, A.},
year={2021},
journal={Patterns},
publisher={Elsevier},
pages={2666-3899}
}

@incollection{Ovchinnikov_2016_1,
author = "Ovchinnikov, I.V.",
title = "Supersymmetric Theory of Stochastics: Demystification of Self-Organized Criticality",
year = "2016",
booktitle = "Handbook of applications of chaos theory",
editor = "Skiadas, C. H. and Skiadas, C.",
publisher = "Taylor\& Francis/CRC",
address = "Boca Raton, London, New York",
}

@BOOK{Gilmore_Lefranc_2002_book,
author = {Gilmore, R. and Lefranc, M.},
title = {The Topology of Chaos: Alice in Stretch and Squeeze land},
publisher = {Wiley},
year = {2002},
address = {New York},
}

@article{Meisel_2012_seizures,
title = {Failure of adaptive self-organized criticality during epileptic seizure attacks},
author={Meisel, C. and Storch, A. and Hallmeyer-Elgner, S. and Bullmore, E. and Gross, T.},
journal={PLoS Comput. Biol. },
year = {2012}, 
volume={8}, 
pages={e1002312},
}

@article{Stoilova_2010_Clebsch_Gordan,
  author  = {Stoilova, N. I. and Van der Jeugt, J.},
  title   = {Gelfand-Zetlin basis and Clebsch-Gordan coefficients for covariant representations of the Lie superalgebra gl(m|n)},
  journal = {Journal of Mathematical Physics},
  volume  = {51},
  number  = {9},
  pages   = {093523},
  year    = {2010},
  month   = sep
}

@article{Candu_2010,
   title={Cohomological reduction of sigma models},
   volume={2010},
   number={5},
   journal={Journal of High Energy Physics},
   publisher={Springer Science and Business Media LLC},
   author={Candu, C. and Creutzig, T. and Mitev, V. and Schomerus, V.},
   year={2010}
}

@article{Quella_2013,
  author       = {Quella, Thomas and Schomerus, Volker},
  title        = {Superspace Conformal Field Theory},
  journal      = {J. Phys. A},
  volume       = {46},
  pages        = {494010},
  year         = {2013},
  doi          = {10.1088/1751-8113/46/49/494010},
  eprint       = {1307.7724},
  archivePrefix= {arXiv},
  primaryClass = {hep-th}
}

@article{Rozansky_1992,
  author  = {Rozansky, L. and Saleur, H.},
  title   = {Quantum Field Theory for the Multi-Variable Alexander-Conway Polynomial},
  journal = {Nuclear Physics B},
  volume  = {376},
  pages   = {461--509},
  year    = {1992}
}

@Article{Watkins_2016,
author="Watkins, N. W.
and Pruessner, G.
and Chapman, S. C.
and Crosby, N. B.
and Jensen, H. J.",
title="25 Years of Self-organized Criticality: Concepts and Controversies",
journal="Space Science Reviews",
year="2016",
volume="198",
number="1",
pages="3--44",
}

@article{Bedard_2006,
  title = {Does the $1/f$ Frequency Scaling of Brain Signals Reflect Self-Organized Critical States?},
  author = {B\'edard, C. and Kr\"oger, H. and Destexhe, A.},
  journal = {Phys. Rev. Lett.},
  volume = {97},
  issue = {11},
  pages = {118102},
  numpages = {4},
  year = {2006},
  month = {Sep},
  publisher = {American Physical Society},
}

@article{Pettersen_2014,
    author = {Pettersen, Klas H. AND Lindén, Henrik AND Tetzlaff, Tom AND Einevoll, Gaute T.},
    journal = {PLOS Computational Biology},
    publisher = {Public Library of Science},
    title = {Power Laws from Linear Neuronal Cable Theory: Power Spectral Densities of the Soma Potential, Soma Membrane Current and Single-Neuron Contribution to the EEG},
    year = {2014},
    month = {11},
    volume = {10},
    pages = {1-26},
    number = {11},
}

@article{PhysRevB.54.1234,
  title = {Charge solitons in one-dimensional arrays of serially coupled Josephson junctions},
  author = {Hermon, Z. and Ben-Jacob, E. and Sch\"on, G.},
  journal = {Phys. Rev. B},
  volume = {54},
  issue = {2},
  pages = {1234--1245},
  numpages = {0},
  year = {1996},
  month = {Jul},
  publisher = {American Physical Society},
  doi = {https://doi.org/10.1103/PhysRevB.54.1234},
}

@article{sine_Gordon_Kink_Bath,
  title = {Overdamped sine-Gordon kink in a thermal bath},
  author = {Quintero, N. R. and S\'anchez, A. and Mertens, F. G.},
  journal = {Phys. Rev. E},
  volume = {60},
  issue = {1},
  pages = {222--230},
  numpages = {0},
  year = {1999},
  month = {Jul},
  publisher = {American Physical Society},
  doi={https://doi.org/10.1103/PhysRevE.60.222}
}

@article{Classics,
Author = {Eilenberger, G.},
Title = {BREMSSTRAHLUNG FROM SOLITONS},
Journal = {Zeitschrift fur Physik B - Condensed Matter},
Year = {1977},
Volume = {27},
Number = {2},
Pages = {199-203},
Publisher = {SPRINGER},
doi={https://doi.org/10.1007/BF01313609}
}

@article{Review_Frenkel_Kontorova,
title = "Nonlinear dynamics of the Frenkel–Kontorova model ",
journal = "Physics Reports",
volume = "306",
number = "1–2",
pages = "1 - 108",
year = "1998",
author = "Braun, O.M. and Kivshar, Y.S.",
}

@article{Josephson_general,
  title = {Soliton excitations in Josephson tunnel junctions},
  author = {Lomdahl, P. S. and Soerensen, O. H. and Christiansen, P. L.},
  journal = {Phys. Rev. B},
  volume = {25},
  issue = {9},
  pages = {5737--5748},
  numpages = {0},
  year = {1982},
  month = {May},
  publisher = {American Physical Society},
  doi = {https://doi.org/10.1103/PhysRevB.25.5737},
}

@Article{TFT_BOOK,
	Title                    = {Topological field theory},
	Author                   = {Birmingham, D. and Blau, M. and Rakowski, M. and Thompson,G.},
	Journal                  = {Phys. Rep.},
	Year                     = {1991},
	Pages                    = {129},
	Volume                   = {209}
}

@article{Witten_1982,
author = {Witten, E.},
title = {Supersymmetry and Morse theory},
journal = {J. Differ. Geom.},
year = {1982},
volume = {17},
pages = {661--692}
}

@article{Slavik_2013,
title = "Generalized differential equations: Differentiability of solutions with respect to initial conditions and parameters ",
journal = "Journal of Mathematical Analysis and Applications ",
volume = "402",
number = "1",
pages = "261 - 274",
year = "2013",
author = "Slavik, A.",
doi={https://doi.org/10.1016/j.jmaa.2013.01.027}
}

@article{Pesin_1977,
author = {Pesin, Ya.},
title = {Characteristic Lyapunov exponents and smooth ergodic theory},
journal = {Russ. Math. Surveys},
volume  = {32},
year = {1977},
pages = {55-114},
doi={https://doi.org/10.1070/rm1977v032n04abeh001639}
}

@Inbook{Arnold_1988,
author="Arnold, L.",
editor="Ziegler, F. and Schu{\"e}ller, G. I.",
title="Lyapunov Exponents of Nonlinear Stochastic Systems",
bookTitle="Nonlinear Stochastic Dynamic Engineering Systems: IUTAM Symposium Innsbruck Igls, Austria, June 21--26, 1987",
year="1988",
publisher="Springer Berlin Heidelberg",
address="Berlin, Heidelberg",
pages="181--201",
}

@article{Ovchinnikov_2020,
title = {Criticality or Supersymmetry Breaking?},
journal = {Symmetry},
volume = {12},
number = {5},
pages = {805},
year = {2020},
author = {Ovchinnikov, I. V. and Li, W. and Sun, Y. and Hudson, A. E. and Meier, K. and Schwartz, R. N. and Wang, K. L.},
doi={https://doi.org/10.3390/sym12050805}
}

@book{Devaney_1992_book,
author = {Devaney, R.},
title = {A First Course in Chaotic Dynamical Systems: Theory and Experiment},
publisher = {Addison-Wesley},
year = {1992}
}

@article{Graham_1988,
author = {Graham, R.},
year = {1988},
journal = {EPL},
volume = {5},
pages = {101},
title = {Lyapunov Exponents and Supersymmetry of Stochastic Dynamical Systems},
doi={https://doi.org/10.1209/0295-5075/5/2/002}
}

@Article{Elaskar_2023,
AUTHOR = {Elaskar, S. and del Río, E.},
TITLE = {Review of Chaotic Intermittency},
JOURNAL = {Symmetry},
VOLUME = {15},
YEAR = {2023},
NUMBER = {6},
ARTICLE-NUMBER = {1195}
}

@incollection{Crutchfield_2003,
author = {Crutchfield, J. P.},
title = {What Lies Between Order and Chaos?},
editor = {Casti, J. and Karlqvist, A.},
booktitle = {Art and Complexity},
publisher = {JAI},
address = {Amsterdam},
pages = {31-45},
year = {2003},
doi={https://doi.org/10.1016/B978-044450944-4/50004-9}
}

@article{Crutchfield_2012,
author = {Crutchfield, J.P.},
year={2012},
journal={Nature Physics},
title={Between order and chaos},
pages={17–24},
volume={8},
doi={https://doi.org/10.1038/nphys2190}
}

@article{Mitchell_1993,
  author  = {Mitchell, M. and Crutchfield, J. P. and Hraber, P.T. },
  title   = {Dynamics, Computation, and the ``Edge of Chaos'': A Re-examination},
  journal = {Complex Systems},
  volume  = {7},
  pages   = {89--130},
  year    = {1993}
}

@book{Kauffman_1993,
  author    = {Stuart A. Kauffman},
  title     = {The Origins of Order: Self-Organization and Selection in Evolution},
  publisher = {Oxford University Press},
  year      = {1993},
  address   = {New York}
}

@article{Langton_1990,
  author  = {Langton, C. G. },
  title   = {Computation at the edge of chaos: Phase transitions and emergent computation},
  journal = {Physica D: Nonlinear Phenomena},
  volume  = {42},
  number  = {1--3},
  pages   = {12--37},
  year    = {1990},
  doi     = {10.1016/0167-2789(90)90064-V}
}

@article{Mangiarotti_2021,
    author = {Mangiarotti, S. and Letellier, C.},
    title = "{Topological characterization of toroidal chaos: A branched manifold for the Deng toroidal attractor}",
    journal = {Chaos: An Interdisciplinary Journal of Nonlinear Science},
    volume = {31},
    number = {1},
    pages = {013129},
    year = {2021},
    month = {01}
}

@book{Hasselblatt_Katok_2002_book,
author={B. Hasselblatt and A. Katok},
title={Handbook of Dynamical Systems},
volume={1A},
year={2002},
publisher={North–Holland, Amsterdam},
}

@incollection{Le_Jan_Watanabe_1984,
title = {Stochastic Flows of Diffeomorphisms},
editor = {Itô, K.},
series = {North-Holland Mathematical Library},
publisher = {Elsevier},
volume = {32},
pages = {307-332},
year = {1984},
booktitle = {Stochastic Analysis},
author = {Le Jan, Y. and Watanabe, S.},
}

@article{Yorke_1975,
author = {Li, T.-Y. and Yorke, J. A. },
title = {Period Three Implies Chaos},
journal = {The American Mathematical Monthly},
volume = {82},
number = {10},
pages = {985-992},
year = {1975},
publisher = {Taylor & Francis}
}

@article{Heirer_2018,
author={Hairer, M.},
title={Renormalisation of parabolic stochastic PDEs},
journal={Jpn. J. Math.},
volume={13},
pages={187–233},
year={2018}
}

@incollection{Arnold_1983,
title = {Qualitative Theory of Stochastic Systems},
editor = {Bharucha-Reid, A.T.},
booktitle = {Probabilistic Analysis and Related Topics},
publisher = {Academic Press},
pages = {1-79},
year = {1983},
author = {Arnold, L. and Kliemann, W.}
}

@InProceedings{Baxendale_1986,
author="Baxendale, P. H.",
editor="Arnold, L. and Wihstutz, V.",
title="The Lyapunov spectrum of a stochastic flow of diffeomorphisms",
booktitle="Lyapunov Exponents",
year="1986",
publisher="Springer Berlin Heidelberg",
address="Berlin, Heidelberg",
pages="322--337",
isbn="978-3-540-39795-3"
}

@InProceedings{Arnold_1986,
author="Arnold, L. and Kliemann, W. and Oeljeklaus, E.",
editor="Arnold, Ludwig and Wihstutz, Volker",
title="Lyapunov exponents of linear stochastic systems",
booktitle="Lyapunov Exponents",
year="1986",
publisher="Springer Berlin Heidelberg",
address="Berlin, Heidelberg",
pages="85--125"
}

@artile{Bedrossian_2022_LyapunovExpReview,
author={Bedrossian, J. and Blumenthal, A. and Punshon-Smith, S.},
title={A regularity method for lower bounds on the Lyapunov exponent for stochastic differential equations.},
journal={Invent. math.},
volume={227}, 
page={429–516},
year={2022}
}

@book{Strogatz_2017_book,
author={Strogatz, S.H.},
year={2015},
title={Nonlinear Dynamics and Chaos: With Applications to Physics, Biology, Chemistry, and Engineering (2nd ed.).},
publisher={CRC Press.}
}

@article{Tel_2015,
    author = {Tél, T.},
    title = {The joy of transient chaos},
    journal = {Chaos: An Interdisciplinary Journal of Nonlinear Science},
    volume = {25},
    number = {9},
    pages = {097619},
    year = {2015},
    month = {07}
}

@article{Szendro_2001,
author={Szendro, P. and Vincze, G. and Szasz, A.},
year={2001},
title={Pink-noise behaviour of biosystems},
journal={European Biophysics Journal},
pages={227-231},
volume={30}
}

@INPROCEEDINGS{Voss_1979,
  author={Voss, R.F.},
  booktitle={33rd Annual Symposium on Frequency Control}, 
  title={1/f (Flicker) Noise: A Brief Review}, 
  year={1979},
  volume={},
  number={},
  pages={40-46}
}

@article{Dutta_1981,
  title = {Low-frequency fluctuations in solids: 1/f noise},
  author = {Dutta, P. and Horn, P. M.},
  journal = {Rev. Mod. Phys.},
  volume = {53},
  issue = {3},
  pages = {497--516},
  numpages = {0},
  year = {1981},
  month = {Jul},
  publisher = {American Physical Society}
}

@ARTICLE{Keshner_1982,
  author={Keshner, M.S.},
  journal={Proceedings of the IEEE}, 
  title={1/f noise}, 
  year={1982},
  volume={70},
  number={3},
  pages={212-218}
}

@article{Ovchinnikov_2024,
title = {Ubiquitous order known as chaos},
journal = {Chaos, Solitons \& Fractals},
volume = {181},
pages = {114611},
year = {2024},
author = {Ovchinnikov, I.V.}
}

@book{RFT_SSB_book,
author = {Brauner, Tom\'as},
title = {Effective Field Theory for Spontaneously Broken Symmetry (Lecture Notes in Physics, 1023)},
publisher = {Springer},
address = {Stavanger, Norway},
year = {2024}
}

@article{Schomerus_Saleur_2006,
title = {The GL(1$|$1) WZW-model: From supergeometry to logarithmic CFT},
journal = {Nuclear Physics B},
volume = {734},
number = {3},
pages = {221-245},
year = {2006},
author = {Schomerus, V. and Saleur, H.}
}

@article{Lee_2010,
author={Lee, K.-M. and Lee, S.-J.},
year={2010},
title={1/2-BPS Wilson loops and vortices in ABJM model},
journal={Journal of High Energy Physics},
issue={9}
}

@article{Maldacena_1998,
  title = {Wilson Loops in Large N Field Theories},
  author = {Maldacena, J.},
  journal = {Phys. Rev. Lett.},
  volume = {80},
  year = {1998}
}

@misc{Natsuume_2016_book,
      title={AdS/CFT Duality User Guide}, 
      author={Natsuume, M.},
      year={2016},
      eprint={1409.3575},
      archivePrefix={arXiv},
      primaryClass={hep-th}
}

@inproceedings{Penedones_2016,
   title={TASI Lectures on AdS/CFT},
   booktitle={New Frontiers in Fields and Strings},
   publisher={World Scientific},
   author={Penedones, Joào},
   year={2016}
}

@article{Gubser_1998,
title = {Gauge theory correlators from non-critical string theory},
journal = {Physics Letters B},
volume = {428},
pages = {105-114},
year = {1998},
author = {Gubser, S.S. and Klebanov, I.R. and Polyakov, A.M.}
}

@article{Witten_1998,
  title={Anti-de Sitter space and holography},
  author={Witten, E.},
  journal={Advances in Theoretical and Mathematical Physics},
  year={1998},
  volume={2},
  pages={253-291}
}

@misc{Kaplan_2016_book,
author={Kaplan, Jared},
title={Lectures on AdS/CFT from the Bottom Up},
url={https://sites.krieger.jhu.edu/jared-kaplan/files/2016/05/AdSCFTCourseNotesCurrentPublic.pdf}
}

@artilce{Wang_2010,
title={Effect of Spatial Charge Inhomogeneity on 1/f Noise Behavior in Graphene},
author={Xu, Guangyu and Torres, Carlos M. Jr. and Zhang, Yuegang and Liu, Fei and Song, Emil B. and Wang, Minsheng and Zhou, Yi and Zeng, Caifu and Wang, Kang L.},
year={2010},
journal={Nano Letters},
volume={10},
pages={3312}
}

@article{Ovchinnikov_2025,
year = {2025},
publisher = {IOP Publishing},
volume = {100},
pages = {125233},
author = {Ovchinnikov, I. V},
title = {Chaos, Ito-Stratonovich dilemma, and topological supersymmetry},
journal = {Physica Scripta}
}

@article{Ovchinnikov_2026,
title = {The supersymmetric origin of chaos and its hidden topological order},
journal = {Newton},
volume = {2},
pages = {100465},
year = {2026},
author = {I. V. Ovchinnikov and M. {Di Ventra}}
}

@book{Altland_Simons_book_2010,
  author    = {Altland, A. and Simons, B.},
  title     = {Condensed Matter Field Theory},
  edition   = {2},
  publisher = {Cambridge University Press},
  year      = {2010},
  doi       = {10.1017/CBO9780511789984}
}

@article{Qi_Zhang_2011,
  author  = {Qi, X.-L. and Zhang, S.-C.},
  title   = {Topological insulators and superconductors},
  journal = {Reviews of Modern Physics},
  volume  = {83},
  number  = {4},
  pages   = {1057--1110},
  year    = {2011}
}

@article{Hansson_2004,
  author  = {Hansson, T. H. and Oganesyan, V. and Sondhi, S. L.},
  title   = {Superconductors are topologically ordered},
  journal = {Annals of Physics},
  volume  = {313},
  number  = {2},
  pages   = {497--538},
  year    = {2004}
}

@misc{Nicolis_2026,
      title={The stochastic approach for anomalies in supersymmetric theories}, 
      author={Nicolis, S.},
      year={2026},
      eprint={2603.29816},
      archivePrefix={arXiv},
      primaryClass={hep-th},
      url={https://arxiv.org/abs/2603.29816}, 
}

@misc{Sethi_Weiderpass_2026,
      title={Supersymmetry and Nonreciprocity}, 
      author={Sethi, S. and Weiderpass, G. A.},
      year={2026},
      eprint={2602.16824},
      archivePrefix={arXiv},
      primaryClass={hep-th},
      url={https://arxiv.org/abs/2602.16824}, 
}

@article{Anselmi_1997,
year = {1997},
volume = {14},
number = {1},
pages = {1},
author = {Anselmi, D.},
title = {Topological field theory and physics},
journal = {Classical and Quantum Gravity}
}

@article{Baulieu_1989,
title = {Stochastic and topological gauge theories},
journal = {Physics Letters B},
volume = {232},
number = {4},
pages = {479-485},
year = {1989},
issn = {0370-2693},
author = {Baulieu, L.}
}

@book{Peskin_Schroeder_1995_book,
  author    = {Peskin, M.E. and Schroeder, D.V.},
  title     = {An Introduction to Quantum Field Theory},
  publisher = {Westview Press},
  year      = {1995},
  address   = {Boulder, CO}
}

@book{Henneaux_Teitelboim_1992_book,
  author    = {Henneaux, M. and Teitelboim, C.},
  title     = {Quantization of Gauge Systems},
  publisher = {Princeton University Press},
  year      = {1992},
  address   = {Princeton},
  isbn      = {9780691024341}
}

@article{Kaviraj_2022,
  title = {Parisi-Sourlas Supersymmetry in Random Field Models},
  author = {Kaviraj, A. and Rychkov, S. and Trevisani, E.},
  journal = {Phys. Rev. Lett.},
  volume = {129},
  issue = {4},
  pages = {045701},
  numpages = {7},
  year = {2022},
  month = {Jul},
  publisher = {American Physical Society}
}

@article{Nakayama_2025,
  title = {Parisi-Sourlas Supertranslation and Scale Invariance without Conformal Symmetry},
  author = {Nakayama, Yu},
  journal = {Phys. Rev. Lett.},
  volume = {134},
  issue = {13},
  pages = {131602},
  numpages = {6},
  year = {2025},
  month = {Apr},
  publisher = {American Physical Society}
}

@book{Rychkov_2023,
  author    = {Rychkov, S.},
  title     = {Lectures on the Random Field Ising Model: From Parisi-Sourlas Supersymmetry to Dimensional Reduction},
  series    = {SpringerBriefs in Physics},
  publisher = {Springer Nature Switzerland},
  year      = {2023},
  isbn      = {978-3-031-41999-7},
  doi       = {10.1007/978-3-031-42000-9}
}

@misc{Le_Floch_2025,
      title={Unleash $Q$! Cohomology, Localization, and Interpolation in Parisi-Sourlas Supersymmetry}, 
      author={Le Floch, B. and Patashuri, G. and Trevisani, E.},
      year={2025},
      eprint={2512.15899},
      archivePrefix={arXiv},
      primaryClass={hep-th},
      url={https://arxiv.org/abs/2512.15899}
}

@article{Kaviraj_2020,
  author        = {Kaviraj, A. and Rychkov, S. and Trevisani, E.},
  title         = {Random Field Ising Model and Parisi-Sourlas supersymmetry. Part I. Supersymmetric CFT},
  journal       = {JHEP},
  year          = {2020},
  volume        = {04},
  pages         = {090},
  doi           = {10.1007/JHEP04(2020)090}
}

@article{Kaviraj_2021,
  author        = {Kaviraj, A. and Rychkov, S. and Trevisani, E.},
  title         = {Random Field Ising Model and Parisi-Sourlas Supersymmetry. Part II. Renormalization Group},
  journal       = {JHEP},
  year          = {2021},
  volume        = {03},
  pages         = {219},
  doi           = {10.1007/JHEP03(2021)219}
}

@article{Cardy_1985,
title = {Nonperturbative aspects of supersymmetry in statistical mechanics},
journal = {Physica D: Nonlinear Phenomena},
volume = {15},
number = {1},
pages = {123-128},
year = {1985},
author = {Cardy, J. L.}
}

@book{LabastidaMarino_2005_book_TQFT,
  author    = {Labastida, J. M. F.  and Mari{\~n}o, M.},
  title     = {Topological Quantum Field Theory and Four Manifolds},
  series    = {Mathematical Physics Studies},
  volume    = {25},
  publisher = {Springer},
  address   = {Dordrecht},
  year      = {2005}
}

@book{Schwarz_1993_book_QFTTopology,
  author    = {Schwarz, A. S.},
  title     = {Quantum Field Theory and Topology},
  publisher = {Springer},
  address   = {Berlin},
  year      = {1993}
}

@article{Maldacena_1999,
author={Maldacena, J.},
year={1999},
title={The large-N limit of superconformal field theories and supergravity},
journal={International Journal of theoretical physics},
volume={38},
pages={1113-1133}
}

@misc{Sadovski_2025_TFTs,
      title={Four-dimensional topological Yang-Mills-Higgs theories with BRST instability}, 
      author={Sadovski, G.},
      year={2025},
      eprint={2510.22323},
      archivePrefix={arXiv},
      primaryClass={hep-th}
}

@misc{Ren_2010,
  author = {Ren, J.},
  title = {One-dimensional holographic superconductor from $AdS_3/CFT_2$ correspondence},
  eprint = {1008.3904},
  archivePrefix = {arXiv},
  primaryClass = {hep-th},
  year = {2010}
}

@article{Alkac_2017,
  author = {Alkac, G. and Chakrabortty, S. and Chaturvedi, P.},
  title = {Holographic P-wave superconductors in 1+1 dimensions},
  journal = {Phys. Rev. D},
  volume = {96},
  number = {8},
  pages = {086001},
  year = {2017}
}

@article{Ker_nen_2009,
   title={Dark solitons in holographic superfluids},
   volume={80},
   journal={Physical Review D},
   author={Ker\"anen, V. and Keski-Vakkuri, E. and Nowling, S. and Yogendran, K. P.},
   year={2009}
}

@article{Ker_nen_2010,
   title={Inhomogeneous structures in holographic superfluids. I. Dark solitons},
   volume={81},
   number={12},
   journal={Physical Review D},
   author={Ker\"anen, Ville and Keski-Vakkuri, Esko and Nowling, Sean and Yogendran, K. P.},
   year={2010}
}

@article{Polyakov_1993,
title = {The theory of turbulence in two dimensions},
journal = {Nuclear Physics B},
volume = {396},
pages = {367-385},
year = {1993},
author = {Polyakov, A. M.}
}

@article{Faddeev_1993,
  author  = {Faddeev, L. D. and Kashaev, R. M.},
  title   = {Topological Quantum Mechanics},
  journal = {Theoretical and Mathematical Physics},
  volume  = {94},
  number  = {2},
  pages   = {166--172},
  year    = {1993}
}

@book{Freed_1999,
  author    = {Freed, D. S.},
  title     = {Five Lectures on Supersymmetry},
  series     = {American Mathematical Society},
  volume     = {32},
  publisher = {American Mathematical Society},
  address   = {Providence, RI},
  year      = {1999}
}

@misc{labastida_1997_lectures,
      title={Lectures in Topological Quantum Field Theory}, 
      author={Labastida, J. M. F.  and Lozano, C.},
      year={1997},
      eprint={hep-th/9709192},
      archivePrefix={arXiv},
      primaryClass={hep-th},
      url={https://arxiv.org/abs/hep-th/9709192}, 
}

@article{Sethi_1994,
title = {Supermanifolds, rigid manifolds and mirror symmetry},
journal = {Nuclear Physics B},
volume = {430},
number = {1},
pages = {31-50},
year = {1994},
author = {Sethi, S.}
}

\end{document}